\documentclass{svjour2}                    
\usepackage{amsmath,graphicx,amssymb,amsfonts}
\begin{document}

\title{Granular Solid Hydrodynamics  (GSH):}
\subtitle{a broad-ranged macroscopic theory of granular media}


\author{Yimin Jiang      \and
        Mario Liu}


\institute{Yimin Jiang \at
             Central South University, Changsha 410083, China
           \and
           Mario Liu \at
             Theoretische Physik, Universit\"{a}t T\"{u}bingen,72076
T\"{u}bingen, Germany \\
              \email{mliu@uni-tuebingen.de} }

\date{Received: date / Accepted: date}

\maketitle
\begin{abstract}
A unified continuum-mechanical theory has been until now lacking for granular media, some believe it could not exist.  
Derived employing the hydrodynamic approach, {\sc gsh} is such a theory, though as yet a qualitative one. The behavior  being accounted for includes static stress distribution, elastic wave, elasto-plastic motion, the critical state and rapid dense flow. The equations and application to  a few typical experiments are presented here.
\end{abstract}
\tableofcontents

\section{Introduction\label{intro}} 
%
The {\it hydrodynamic formalism}  was pioneered by Landau~\cite{LL6} and Khalatnikov~\cite{Khal} in the context of superfluid helium, and introduced to complex fluids by de Gennes~\cite{deGennes}. (Most physicists take {\em hydrodynamics} to mean the long-wave-length continuum theory of any condensed system, while engineers typically use it as a synonym for the Navier-Stokes equations.) The formalism considers energy and momentum conservation simultaneously, and has a tailored set of state variables for each condensed system. In contrast, the usual approach via {\it constitutive relations} typically leaves out energy conservation and considers the same set of  variables for all systems.  

In deriving a constitutive relation for a complex fluid, one usually focuses on its  rheology, and postulates a quantity $\mathfrak{C}_{ij}$, as a function of the  stress $\sigma _{k\ell}$, strain rate $v_{mn}$, and density $\rho $, such that the constitutive relation ${\partial}_{t}\sigma _{ij}=\mathfrak{C}_{ij}$ holds. (${\partial}_{t}\equiv\partial/\partial t$ needs to be replaced by an objective derivative more generally.)  
Together with the continuity equation $\partial _{t}\rho +\nabla _{i}(\rho v_{i})=0$, momentum conservation, $\partial _{t}(\rho v_{i})+\nabla_{j}(\sigma _{ij}+\rho v_iv_j)=0$, it forms a closed set of equations for $\rho $, $\sigma _{k\ell}$ and the velocity $v_{i}$, which one may take as the universal set of variables. 
The function $\mathfrak{C}_{ij}$ is specified employing experimental data, of which, usually, only a subset is employed, say elasto-plastic motion in granular media, but then not fast dense flow or elastic waves. Even though, in complex systems, it is a difficult approach requiring many arbitrary steps. 

This is far better in the  hydrodynamic approach.  
In deriving a theory, one first identifies the basic physics of a system, with the help of which a set of state variables is specified. Different complex systems with different underlying physics therefore have different state variables. Then, by considering energy and momentum conservation, in addition to entropy balance, the energy flux and the stress are derived (not postulated), as functions of the state variables and their spatial derivatives, with a clear distinction between dissipative and reactive terms. Given the stringency of derivation, therefore, the  continuum-mechanical theory thus obtained is the appropriate one for the given system if the input in physics is adequate. 

Being a subject of practical importance, elasto-plastic deformation of dense granular media has been under the focus of engineering research for many decades if not
centuries~\cite{schofield,nedderman,wood1990,kolymbas1,kolymbas2,gudehus2010}. 
The state of the art, however, is confusing: A large number of constitutive models compete, employing strikingly different
expressions, with none accepted as authoritative. In his recent book, {\it Physical Soil Mechanics}~\cite{gudehus2010},  Gudehus uses phrases such as {\em morass of equations} and {\em jungle of data} as metaphors. 
Moreover, this competition is among theories applicable only to elasto-plastic deformation, while rapid dense flow is taken to obey yet rather different equations~\cite{hutter2007}. 

It took us a while to understand, but now we realize that although these theories achieve considerable realism,  they are in essence clever renditions of complex data, not reflections of the underlying physics. 
This is the reason it appears worthwhile to us trying out the hydrodynamic approach, by focusing on the physics first, leaving the rich and subtle granular phenomenology aside while constructing the theory. Our hope is to  arrive at one that, though not necessarily accurate in every aspect, is  firmly based in physics, applicable over the complete range of shear rates,  and affords a well founded understanding. 

Hydrodynamic theories~\cite{hydro-1,hydro-2} have been derived for many condensed systems, including liquid crystals
~\cite{liqCryst-1,liqCryst-2,liqCryst-3,liqCryst-4,liqCryst-5,liqCryst-6,liqCryst-7},
superfluid $^3$He~\cite{he3-1,he3-2,he3-3,he3-4,he3-5,he3-6},
superconductors~\cite{SC-1,SC-2,SC-3}, macroscopic
electro-magnetism~\cite{hymax-1,hymax-2,hymax-3,hymax-4},
ferrofluids~\cite{FF-1,FF-2,FF-3,FF-4,FF-5,FF-6,FF-7,FF-8,FF-9}, and
polymers~\cite{polymer-1,polymer-2,polymer-3,polymer-4}. It is useful and possible also for granular media: Useful, because it should help to illuminate and order their
complex behavior; possible, because a significant portion has 
already been done. We call it {\sc gsh}, for ``{\it granular solid hydrodynamics}.''
Remarkably, it divides granular behavior into three regimes, with the granular temperature $T_g$, or  jiggling of the grains, serving as a dial:

\begin{enumerate}
\item At $T_g\to0$, grains hardly jiggle, implying vanishing shear rates $\dot\gamma$: Static stress distribution and the propagation of elastic waves are phenomena of this regime. We call it {\em quasi-elastic} because the stress stems from deformed grains and is elastic in origin, and because the terms responsible for  plastic behavior are quadratically small. 
\item At slightly elevated $T_g$ and slow rates: The 
stress is still predominantly elastic, but it may now relax: When the grains loose (or loosen) contact with one another, both granular deformation and the associated stress will decrease. The plastic terms are now comparable to the elastic ones. Typical phenomena are the {\em critical state}~\cite{schofield} and {\em incremental nonlinearity} (ie. the strikingly different loading and unloading curves), not seen in the quasi-elastic regime. The hypoplastic model~\cite{kolymbas1} and other rate-independent constitutive relations hold here. We call it the {\em hypoplastic regime} -- without implying the lack of potentials as originally thought.
\item At large $T_g$ and high shear rates, we have the rapid dense flow behavior covered by the $\mu(I)$-model~\cite{midi} and Bagnold flow. The jiggling is so strong that it exerts a pressure, and viscosities are important. They compete with the elastic stress, becoming dominant at  high rates and low densities.  
\end{enumerate}

Finally, some words on the difference between the structure and parameters of a theory: The first  concerns the part that  is derived from general principles, the second is a material-dependent input, typically an assumption. This difference is especially clear-cut in constitutive models, where the structure is given by the laws of mass and momentum conservation, and the parameter is given by $\mathfrak{C}_{ij}$. In a hydrodynamic theory, the structure consists of (1)~the conservation laws for energy, momentum and mass, and (2) their respective fluxes, given as functions of the state and conjugate variables (eg. entropy $s$ and temperature $T\equiv\partial w/\partial s$), including their spacial derivatives (eg. $\nabla_iT$). In this form, the fluxes are generally valid. The most important parameter is the expressions for the energy $w$. Once given,  the conjugate variables, and with them also the fluxes, are explicit functions of the state variables. Transport coefficients (eg. the viscosity) are also parameters.   Ideally, one would like to obtained the parameters from a microscopic theory, though this is a tall order  accomplished mainly in dilute systems (or formally dilute ones). 
For denser ones, the realistic approach is to engage in a trial-and-error iteration, in which the ramification of postulated dependences are compared to experiments and simulations. 

The structure of {\sc gsh} is, we believe, complete and adequate, but the parameters are not as yet specified with  complete confidence. The main reason is, although the dependency of the transport coefficients on the granular temperature $T_g$ seems fairly universal, obtainable from more general considerations, that on the density varies with the type of grains. And this system-specific information needs to be obtained from data on only one type of grains. (In a sense, the $T_g$-dependence is  more {\it structural}.)

In what follows, we shall first discuss the basic physics of granular media in Sec~\ref{intro-3}, then present the equations of {\sc gsh} in Sec~\ref{GSH}. The next two sections are respectively devoted to granular behavior in the quasi-elastic and hypoplastic regime, as defined above. (Application of {\sc gsh} to fast dense flow of Regime 3 is in~\cite{p&g2013}.) The manuscript ends with a conclusion and a list of symbols.

\section{The Basic Physics of Granular Media\label{intro-3}} 
\subsection{Two-Stage Irreversibility}
\begin{figure}[b] 
\begin{center}
\vspace{-2cm}
\includegraphics[scale=0.18]{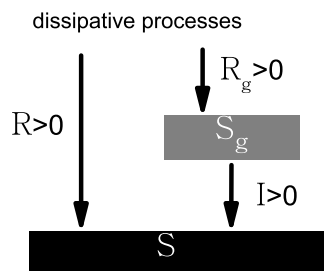}
\end{center}
\vspace{-1.5cm}
\caption{\label{2stageIrr}
{\em Two-stage irreversibility}. Dissipative processeses produce either granular entropy $S_g$, or directly thermal entropy $S$. Eventually, $S_g$ is also converted to $S$.}
\end{figure}

To derive the
hydrodynamic theory for granular media, one needs the input of what the
essence of granular physics is. We believe it is
encapsulated by two notions: {\em two-stage irreversibility} and {\em variable transient elasticity}. The first is related to the three spatial scales of any
granular media: (a)~the macroscopic, (b)~the mesoscopic, granular, and (c)~the
microscopic, inner granular. Dividing all degrees of freedom (DoF) into these three
categories, we treat those of (a) differently from (b,c). Macroscopic
DoF: the slowly varying stress, flow and density fields, are
employed as state variables, but inter- and
inner granular DoF are treated summarily: 
Only their contributions to the energy is considered and taken,
respectively, as granular and true heat. So we do not account for the motion of a jiggling grain, only include its fluctuating kinetic and elastic energy as contributions to the granular 
heat, $\int T_g{\rm d}S_g$, characterized by the granular entropy $S_g$ and temperature $T_g$. Similarly, phonons are taken as part of true heat, $\int T{\rm d}S$. There are a handful of macroscopic DoF~(a), a large number of granular ones (b), and yet many orders of magnitude more inner granular ones (c). So the statistical tendency to equally distribute the energy among all DoF implies an energy decay: (a) $\to$ (b,c) and (b) $\to$ (c). 
This is what we call {\em two-stage irreversibility}, see Fig~\ref{2stageIrr}

The system is in equilibrium if the true entropy $S$ is maximal.
Maximal $S_g$ would characterize a quasi-equilibrium only
if there were no energy decay from (a,b) to (c), or when it is slow enough to be neglected. As the ubiquitous inelasticity of granular collisions  demonstrate, this is never the case. 

A division into three scales works well when they are clearly separated -- though this is a problem of accuracy, not viability. Scale separation is well satisfied in large-scaled,  engineering-type experiments, but less so in small-scaled ones. Using glass or steel beads aggravates the problem. The same is true of 2D experiments employing less and larger disks. On the other hand, when there is too little space for spatial averaging, one may still average over time and runs, to get rid of fluctuations not contained in a hydrodyanmic theory.  

\subsection{Significance and Measurement of $\boldmath{T_g}$\label{significanceT_g}}
Thermodynamic variables are, strictly speaking,  either conserved ones, such as energy and mass, or broken symmetry ones, such as the preferred direction in nematic liquid crystals. But one can conveniently include slowly relaxing variables in an appropriately generalized thermodynamic theory.  A  well known example for such variables is the magnitude of the order parameter $\psi$ in a Ginzburg-Landau theory, such as the superfluid density $\rho_s\sim\psi\psi^*$~\cite{LL6,LL5}.  Expanding the free energy in $\psi$ at the phase transition, one obtains $f=\alpha|\psi|^2+\beta|\psi|^4+\cdots$, with  equilibrium given for minimal $f$, or $\partial f/\partial\psi=0$. Circumstances are especially simple for $\alpha>0$, when the term $\sim\beta$ is negligible, and $\psi=0$ the equilibrium condition. 

The idea behind this is a generalized notion of equilibrium -- call it {\it quasi-equilibrium} -- from one in which  $\psi$ vanishes to one with a given value of $\psi$. The associated statistical ensemble includes only the micro-states compatible with this value. These are not as numerous as those with $\psi=0$, but more than sufficient for the consideration to hold, that quasi-equilibrium is the macro-state with the largest number of compatible micro-states. Any thermodynamic consideration that derives from it remains well founded, especially the principle of maximal entropy. As a result, the conjugate variables retain their thermodynamic significance: For instance, $P\equiv-\left.{\partial W}/{\partial V}\right|_{\psi}$ (with $W$ the energy and $V$ the volume) is the equilibrium force for given $\psi$. 

Conserved and relaxing variables have different equilibrium conditions. The latter assumes a specific value (frequently zero, and more generally given by the vanishing of the conjugate variable, $\partial W/\partial\psi=0$). The former does not, though the associated conjugate variable is constant. For instance, the equilibrium condition with respect to energy exchange between two systems is equal temperatures, $T_1=T_2$, and to mass exchange equal chemical potentials, $\mu_1=\mu_2$. Curiously, $T_g$ alternates between both types of behavior. 
 
Before we enter into its discussion, a caveat and a note. 
Writing $\int T_g{\rm d}S_g$ for the energy contained in the granular DoF is useful only if they are in equilibrium with one another. This may not always be the case, eg. in a granular gas maintained by vibrating walls. (The system therefore needs additional state variables to characterize the velocity distribution, see~\cite{2-p&g2013}.) But grains are increasingly better equilibrated for higher densities and more frequent collisions. To keep the discussion simple, we assume that they are always in equilibrium. The note: Static granular ensembles are always in (a properly understood) thermodynamic equilibrium. Both  the force equilibrium and $T_g=T$ are results of maximizing the total entropy, see Sec~\ref{equicond}.

In a rarefied granular gas, the energy $W$ has only kinetic contributions. Equipartion implies we have $\frac12T_g$ per DoF, or $W=\frac32T_gN$ (with $N$ the number of grains).  
Assuming that the inner granular DoF may be modeled as a phonon gas, we take its energy as  $3TN_a$ (for $T\gg T_D$, the Debye temperature, and $N_a$ the number of atoms in all the grains). If the grains maintained their mechanical integrity at arbitrarily high $T$, they will heat up during a collision for $T_g>T$, but cool down for $T_g<T$ (by amplifying the Brownian motion), until they are in equilibrium, for $T_g=T$. Clearly, all this results from $T_g$ being associated with the conserved energy (or equivalently, with the ``conserved'' entropy while minimizing the energy). Usually, of course, because $T_g\gg T$, the heat transfer is taken as independent of $T$, given by a constant restitution coefficient. It then relaxes, like $\psi$, until it vanishes. 
 
Since the grains collide more frequently at higher densities, increasing the elastic contribution to the energy, this simple picture gets blurred,  breaking down completely when the contacts become enduring. 
Given the complicated interaction between grains at higher densities, including dissipation and friction, a valid $W(S_g)$ with $T_g\equiv\partial W/\partial S_g$ seems difficult to obtain head-on. Therefore, we choose the following pragmatic approach. Starting from the energy density as a function of the two entropy densities, $w(s,s_g)$, we write 
${\rm d}w=T{\rm d}s+T_g{\rm d}s_g=T{\rm d}(s+s_g)+(T_g-T){\rm d}s_g$,
identifying the first term as the equilibrium energy for  $T_g=T$, and the second as the additional contribution  $\Delta w$ if  $T_g\not=T$.  Written this way, $T$ is
associated with the conserved total entropy $s_{tot}\equiv s_g+s$. It does not have a definite equilibrium value, but will equalize with the temperature of another system if heat exchange is allowed. Associated with $s_g$ at given $s_{tot}$, $T_g-T$ is an internal excitation of a  non-optimal energy distribution, akin to mass nonuniformity (that will relax if uniform mass is the equilibrium state, because mass conservation does not come in here). Therefore, $T_g-T$ relaxes until  $T_g-T=0$. 
Using the same arguments as employed for the Ginzburg-Landau free energy, that the energy $\Delta w(\rho,T,T_g-T)$ has a minimum for  $T_g-T=0$,  we expand $\Delta w$ 
to find
\begin{equation}\label{2b-1}
\Delta w={s_g^2}/{2\rho b}, \quad T_g-T\equiv \left.{\partial w}/{\partial s_g}\right|_{s_g+s}={s_g}/{\rho b},
\end{equation}
with $b=b(\rho,T)>0, \Delta w=\frac12\rho b (T_g-T)^2$. 
As this consideration assumes only analyticity of $w$ and  does not depend on the interaction, it is quite general. So Eq~(\ref{2b-1}) should hold for $T_g$ sufficiently small, 
and the remaining question is what $b$ is. 
First, we note that taking the dimension of $T_g, s_g$ as energy and inverse volume, respectively, that of $1/b$ is volume$\times$energy.  Next, having established the quadratic dependence (surprising as we are used to $w\sim T_g$), we may now take the realistic limit, $T_g\gg T$, $s\gg s_g$, to realize that the above rewriting of ${\rm d}w$ did not change much, since $s\approx s_{tot}$, $T_g\approx T_g-T$. But we now  do know that $T$ is associated with a  conserved variable, while $T_g$ is, {\it cum grano salis}, a relaxing one. And we may consider  $b=b(\rho)\equiv b(\rho,T\to0)$.

As remarked at the end of the introduction, density dependence is a system-specific property,  that needs to be obtained from experiments or simulation. We note that a gas of light-weight and  completely elastic beads may serve as a {\it granular thermometer} in a  DEM-experiment for a regular granular system comprising of dissipative, heavy grains, if both are separated by a massless membrane (that will transmit momentum but no particles). Maintaining the regular system at $\rho,\langle v\rangle=$ const,
where $\langle v\rangle^2\equiv\langle\vec v_i\cdot\vec v_i\rangle$, one can measure its energy $w(\rho,\langle v\rangle)$, both the kinetic and the elastic contributions, and read its temperature $T_g(\rho,\langle v\rangle)$ off the thermometer. Combining both yields $w=w(\rho,T_g)$, or $b(\rho)$ via an expansion of $w$ in $T_g$. 

This procedure works mainly because waiting long enough, the thermometer will  equilibrate with the grains, irrespective how dissipative they are.  The many reasons a real experiment would not work is of course related to the conclusion we draw above that $T_g$ is primarily a relaxing quantity that seeks to attain its local equilibrium value, irrespective whether there is another system of a different $T_g$, with which energy may be  exchanged. Given this lack of circumstances in which the thermodynamic significance of $T_g$ plays any robust role, it seems futile to insist on it. 
So, instead of measuring $b$ via $T_g$'s thermodynamic significance, we may  employ Eqs~(\ref{2b-1}), with a postulate $b(\rho)$ [such as given in Eqs~(\ref{2b-5}) below], to define $T_g$, such that it holds for all $T_g,\rho$. Although $T_g$ is then a true temperature only at  $T_g=T$, this strategy works surprisingly well, not only for elasto-plastic motion in dense media, but also for gases. 
In {\sc gsh}, the pressure exerted by jiggling grains is $\sim b T_g^2$, the viscosity $\sim T_g$, see Eq~(\ref{2b-5}, \ref{2c-3}) below. Equating $\Delta w=\frac12\rho bT_g^2$ to $\frac32 T_k\rho/m$ for 
granular gases, we obtain 
\begin{equation}\label{2b-2c}
3T_k/m=b(\rho) T_g^2.
\end{equation}
And indeed, the pressure is found $\sim T_k$, the viscosity $\sim\sqrt{T_k}$,  in an approach combining the kinetic theory and DEM results~\cite{luding2009,Bocquet}. 

Finally, it seems useful to probe whether one may identify $T_g$ with the average velocity $\langle v\rangle$. Taking the energy density of the granular DoF as $w=\rho c(\rho)\langle v\rangle^2/2$, we have $c=1$ in the dilute limit, and may plausibly take $c(\rho)$ approaching 2 in the dense limit, where enduring contacts dominate, hence any kinetic energy is converted into elastic one at the next instance, and back again. However, this conjecture, conveniently linking a macroscopic to a mesoscopic quantity, needs to be thoroughly validated. 

\subsection{Variable Transient Elasticity}
Our second notion, {\em variable transient elasticity}, addresses granular plasticity.
The free surface of a granular system at rest is frequently tilted. When
perturbed, when the grains jiggle and $T_g\not=0$, the tilted surface will
decay and become horizontal. The stronger the grains jiggle
and slide, the faster the decay is. We take this as indicative of a system
that is elastic for $T_g=0$, transiently elastic for $T_g\not=0$,
with a stress relaxation rate $\sim T_g$. 

A relaxing stress is typical of any viscous-elastic system such as polymers~\cite{polymer-1}. 
The unique circumstance here is that the relaxation rate is not a material
constant, but a function of the state variable $T_g$. As we shall see, it
is this {\em variable transient elasticity} -- a simple fact at
heart -- that underlies the complex behavior of granular plasticity.
Realizing it yields a most economic way to capture granular rheology at elasto-plastic rates.

Employing a strain field rather than the
stress as a state variable usually yields a simpler description, because the former is in
essence a geometric quantity, the latter a physical one -- compare the evolution equations for both. Yet one cannot use the
standard strain field $\epsilon_{ij}$ as a granular state variable, because the relation between stress and $\epsilon_{ij}$ lacks uniqueness when the system is plastic.
Engineering theories frequently divide the strain into two fields, elastic $u_{ij}$ and plastic
$\epsilon^{p}_{ij}$, with the first accounting for
the reversible and second for the irreversible part. They then employ $\epsilon_{ij}$ and $\epsilon^{p}_{ij}$ as two independent
strain fields to account for elasto-plastic motion of granular media~\cite{Houlsby,Houlsby2}.
We believe that, on the contrary, the elastic strain $u_{ij}$
is the sole state variable, as there is a unique relation between the elastic stress
$\pi_{ij}$ and $u_{ij}$, as convincingly argued by Rubin~\cite{rubin}. We take $u_{ij}$ as the portion of the strain that deforms the grains and  changes the energy $w=w(u_{ij})$. And since an elastic stress $\pi_{ij}$ only exists when the grains are deformed, it is also a function of $u_{ij}$.
Employing $u_{ij}$ as the sole state variable preserves many useful
features of elasticity, especially the (so-called hyper-elastic) relation,
\begin{equation}\label{1-1} \pi_{ij}=-\partial w(u_{ij})/\partial u_{ij}.
\end{equation} 
This is derived in~\cite{granR2} but easy to understand via
an analogy. The wheels of a car driving up a snowy hill will grip the ground  part of the time, slipping  otherwise. When the wheels grip, the car moves and its gravitational energy $w$ is increased. Dividing the wheel's rotation $\theta$ into a gripping $\theta^{(e)}$ and a slipping $\theta^{(p)}$ portion, we may compute the torque on the wheel as $\partial w/\partial\theta^{(e)}$  (if the wheel turns sufficiently slowly), same as in Eq~(\ref{1-1}). How much the wheel turns or slips, how large $\theta$ or $\theta^{(p)}$ are, is irrelevant for the torque.

Although $\epsilon_{ij}$ is not a state variable, there are effects of the rates that need to be included: Given a shear rate $\dot\gamma$, the grains will jiggle and slide, producing a finite $T_g$.  This effect is included in the balance equation for $T_g$, which accounts for the energy decay from (a) to (b). It has two  steady-state limits, $T_g\sim\dot\gamma^2$ for low shear rates, and  $T_g\sim\dot\gamma$ for higher ones.

The only way to find out whether {\it two-stage irreversibility} and {\it variable transient elasticity} are appropriate and sufficient, is to derived the theory and compare its ramifications with experiments. The structure of the theory (that we call {\sc gsh}) has already been derived, see~\cite{granR2,granR3,gudehus-jl}, though it was written such that the formal derivation is stressed, not the results, making them less accessible. And there were some blanks left, especially the density dependence of the transport coefficients, and the dependence of the elastic energy on the third invariant. All results are presented here in a readable way, and with the blanks filled in as far as possible. The second step, finding the ramifications, is a more lengthy process, in the midst of which we are.

\subsection{Validity of General Principles\label{intro-2}} 
{\sc gsh} is derived employing conventional methods of theoretical physics,  assuming thermodynamic considerations and associated general principles, especially the Onsager relation, are valid in granular media. As some in the community do not subscribe to it (possibly following  Kadanoff~\cite{kadanoff}, who conjectured early on that granular media, being unique, may not have a hydrodynamic theory), we lay out our reasons why we believe granular media are not different to the extend as to actually violate general principles. 

First, we distinguish between a {\it general principle} and an analogy. The first has been proven to hold under general conditions, hence the name; the second may be good or bad, though no basic result of theoretical physics is imperiled whatever its validity. For instance, 
there are two versions of the {\it fluctuation-dissipation theorem} in granular media, one in terms of the true temperature $T$, the other in terms of the granular temperature $T_g$. The former is a {general principle} that is equally applicable to gas, a block of copper and a pile of sand, quantifying how much, eg. the volume of each fluctuates. The latter derives from the  analogy between $T_g$ and $T$, and has been shown to be invalid at times -- hardly surprising,  since the analogy is far from perfect. Similarly, while energy is frequently deemed not to be conserved in granular media, it is in fact only the kinetic energy of the grains that is not conserved. The total energy, including the heat in the grains, of course is.

Then there is the argument~\cite{onsager} that since  grains collide inelastically and execute irreversible motion, and since the validity of the Onsager relation depends on the {\it time reversal invariance} of the underlying microscopic dynamics, the Onsager relation does not hold in granular media. This argument is not convincing -- the fact that {\it granular kinetic theory is irreversible because it is mesoscopic} has been overlooked here. The true microscopic dynamics in sand is, as everywhere else, the reversible Schr\"{o}dinger equation for the constituent atoms. 
It  is a deeply held belief in theoretical physics that all systems obey CPT-invariance.  In condensed matter, with only electromagnetic interaction, T-invariance holds. This is the foundation of the Onsager relation, a general principle, see eg. the proof in~\cite{LL5}. The specificity of the system, or the  theory one happens to employ, are irrelevant for its validity. 


Another argument states that, since a sand pile has much more gravitational energy than a monolayer of grains, only the latter, the minimal energy state, is in equilibrium. The former, being ``{\it jammed}'' and prevented to reach the latter,
is too far off equilibrium for thermodynamics to hold. We contend that one needs to first also include in the consideration the elastic energy; and second, to realize that a stuck piston,
positioned between two chambers of air, is also ``jammed.'' Yet this is a system in equilibrium because all its many degrees of freedom are except one: the position of the piston that upholds a {\it constraint} on the volumes of the two subsystems.   Thermodynamics is routinely applied to such a system.
In a macroscopic body, all elastic DoF are in equilibrium if the force balance holds, implying the sum of gravitational and elastic energy is minimal, see Sec.\ref{equicond}. Two elastic bodies, one on top of another, are also in equilibrium if the sum of their energy is minimal -- though there is the constraint that the upper body must not slide with respect to the lower one. A sand pile is many little elastic bodies on top of one another. If they are constrained to stay put, and their total energy is minimal, the pile is in equilibrium and amenable to thermodynamic considerations.

\section{The Expressions of {GSH} \label{GSH}    } 
The expressions of {\sc gsh} are divided into the static and dynamic parts. 
{\bf Statics} includes the state variables, the formal equilibrium conditions in terms of them, and the expression for the thermodynamic energy. We also discuss the convexity transition of the energy, how it accounts for yield surfaces in the variable space, beyond which no elastic solutions remain stable. We note that it is qualitatively different from the yield-like critical state, one being a static, the other a dynamic, phenomenon.  
{\bf Dynamics} includes conservation laws, balance equations for $s_g$ and $s$, and evolution equations for the rest of the state variables. Explicit expressions for the transport coefficients, the energy flux and the Cauchy stress are given.

\subsection{Granular Statics\label{gsh-sta}} 
\subsubsection{Complete Set of State Variables}
In accordance to the above stated understanding of granular media's basic physics, the state variables are: the granular entropy $s_g$ and the elastic strain $u_{ij}$, in addition to the usual variables: the density $\rho$, the momentum density $\rho v_i$, the true entropy $s$. Denoting the
energy density (in the rest frame, $v_i=0$) as $w=w(\rho, s, s_g, u_{ij})$,
the conjugate variables are: 
\begin{equation}\label{2-2} \mu\equiv\frac{\partial w}{\partial\rho},\quad
T\equiv\frac{\partial w}{\partial s},\quad T_g\equiv\frac{\partial
w}{\partial s_g},\quad \pi_{ij}\equiv-\frac{\partial w}{\partial u_{ij}},
\end{equation} 
where $\mu$ is the  chemical potential, $T$ the temperature, $T_g$ the 
granular temperature, and $\pi_{ij}$ the elastic stress. These are given once the energy $w$ is.
 %
Next, in Sec~\ref{equicond}, equilibrium conditions will be
derived formally, in terms of the energy and its conjugate variables, whatever $w$ is. Then, in Sec~\ref{granEn}, an example for $w$ will be given, and the conjugate variables calculated -- with the help of which the equilibrium conditions are rendered  explicit.

A complete set of state variables is one that determines a
unique macroscopic state of the system. If a set is given, there is no room
for ambiguity, for ``history-" or ``preparation-dependence." Conversely, any such
dependence indicates that the set is incomplete. In the hydrodynamic
approach, a physical quantity is a state variable if (and only if) the
energy $w$ depends on it. Our assumption is that the above set is
complete.


%
\subsubsection{Formal Equilibrium Conditions\label{equicond}} 
Equilibrium conditions for the state variables, usually in terms of
their conjugate variables, are obtained by requiring the entropy $\int s\,{\rm d}^3r$
to be maximal with appropriate constraints: constant energy $\int w\,{\rm d}^3r$, constant mass $\int \rho\, {\rm d}^3r, \cdots$. In granular media, remarkably, this universally
valid procedure leads to two distinct sets of equilibrium conditions, the
solid- and the fluid-like one. Maximizing the entropy (see~\cite{granR2,granR3} for details), we first
obtain the condition of uniform true temperature $\nabla_iT=0$, and the
requirement that the granular temperature vanishes, $T_g=0$. Usually, $T_g$
vanishes quickly, and if it does, the density is not independent from
the elastic strain, $d\rho/\rho=-du_{\ell\ell}$. They share a
common condition that we identify as the solid one,
\begin{eqnarray}
\label{2a-1}
\nabla_i(\pi_{ij}+P_T\delta_{ij})=\rho\, {\rm g}_i,\,\,
\quad 
P_T\equiv-\partial(wV)/\partial V
\stackrel{T_g\to0}{\longrightarrow}0,
\end{eqnarray} 
where ${\rm g}_i$ is the gravitational constant, $\pi_{ij}$ the elastic stress,
$P_T$ the usual expression for the fluid pressure, and $V$ the volume.  (The derivative is taken at constant $\rho V$, $sV$ and  $s_gV$.) 
With the energy expression  $w$ of the next Sec~\ref{granEn}, $P_T\sim T_g^2$ is the pressure exerted by jiggling grains. We therefore call it the {\em seismic pressure}~\cite{gudehus-jl}.  Clearly, equilibrium condition Eq~(\ref{2a-1}), expressing force balance, is logically the result of maximal true entropy. 

If $T_g$ is kept finite by external perturbations, the system may further increase its entropy by independently varying $\rho$ and $u_{ij}$, to arrive at the fluid equilibrium. It is characterized by two conditions, the first with respect to $u_{ij}$, and the second with respect to $\rho$: 
\begin{equation}\label{2a-2} \pi_{ij}=0, \quad
\nabla_iP_T=\rho\, {\rm g}_i. \end{equation} 
The first condition requires shear stresses to vanish in equilibrium, and free surfaces to be horizontal. The second governs reversible compaction.

\subsubsection{Granular Energy\label{granEn}} 

Interested in stiff grains with small $u_{ij}$, we look for the lowest order terms in the elastic energy $w_\Delta$. Denoting $\Delta\equiv -u_{\ell\ell}$, $P_\Delta\equiv\pi_{\ell\ell}/3$, $u_s^2\equiv
u^*_{ij}u^*_{ij}$, $\pi_s^2\equiv \pi^*_{ij}\pi^*_{ij}$, where
$u^*_{ij},\pi^*_{ij}$ are the respective traceless tensors, we take it as
\begin{eqnarray}\label{2b-2}  w=w_T+w_\Delta,\quad 
&w_T&={s_g^2}/({2\rho b}), \quad 
w_\Delta=\sqrt{\Delta }(2 {\mathcal B} \Delta^2/5+ {\mathcal A}u_s^2),
\\\label{2b-2a} 
\pi_{ij}=\sqrt\Delta({\cal B}\Delta&+&{\cal A}
{u_s^2}/{2\Delta})\delta _{ij}-2{\cal A}\sqrt\Delta\, u_{ij}^*, 
\\\label{2b-2b} 
P_\Delta=\sqrt\Delta({\cal B}\Delta&+&{\cal A}
{u_s^2}/{2\Delta}),\quad \pi_s=-2{\cal A}\sqrt\Delta\, u_s. 
\end{eqnarray} 
Note $u_{ij}$ and $\pi_{ij}$ are collinear and have the same principal axes. The contribution 
$w_T$ is an expansion in $s_g$, as discussed in detail around Eq~(\ref{2b-1}).
%
%
Fixing the density-dependence of the coefficient $b$ yields a
contribution for the seismic pressure $P_T\equiv-\partial(wV)/\partial(V)$. (There is also one from $w_\Delta\sim\Delta^{2.5}$ that is always much
smaller than $P_\Delta\sim\Delta^{1.5}$ for small $\Delta$, and hence neglected.) With $\rho_{cp}$ the random close density, we take 
\begin{equation}\label{2b-5}
b=b_0\left(1-\rho/{\rho_{cp}}\right)^a,\quad
P_T={\rho^2\,a b\, T_g^2}/{2(\rho_{cp}-\rho)}, \end{equation}
where $b_0$ and $a$ being positive numbers. Given Eq~(\ref{2b-2c}) (noting the density dependence of $b$), this is essentially the familiar pressure expression $\sim
T_G/(\rho_{cp}-\rho)$, see eg.~\cite{Bocquet}.    
Fast dense flow experiments appear to point to a small $a$, say
$a\approx0.1$, see~\cite{denseFlow}.   (As we are not, at present, interested in effects such as thermal expansion, the dependence on $s$ is not discussed.)

The second term $w_\Delta$ of Eq~(\ref{2b-1}), with ${\cal A,B}>0$, is the elastic contribution. Its order of 2.5 is important for many granular features, especially {\em stress-induced anisotropy} (see below) and the {\em convexity transition}, discussed in the next section,  Sec~\ref{yield surfaces}. The associated stress expression $\pi_{ij}$ has been validated for the following circumstances, achieving good agreement: 
\begin{itemize}
\item Static stress distribution in three classic geometries: silo, sand pile,
point load on a granular sheet, calculated using the equilibrium condition, Eq~(\ref{2a-1}), see~\cite{ge1,ge2}. 
\item Small-amplitude stress-strain relation, see~\cite{kuwano2002,ge3}.
\item Anisotropic propagation of elastic waves, see~\cite{jia2009,ge4}. 
\end{itemize}
An explanation of ``{\it stress-induced anisotropy}'': In  linear elasticity $w\sim u_s^2$, we have constant second derivatives $\partial^2w/\partial u_s^2$, and the velocity of a elastic wave $\sim\sqrt{\partial^2w/\partial u_s^2}\,$ does not depend on the elastic strain, or equivalently, the stress. For any exponent other than 2, the velocity depends on the stress, and is anisotropic if the stress is.  

Note that the energy $w=w_T+w_\Delta$ vanishes when the grains are neither deformed nor jiggling: $w\to0$ for $s_g, u_{ij}\to0$, implying the lack of any longer-ranged
interaction among the grains. If there were one, there would be a density-dependent term in
$w$ that remains finite for  $s_g, u_{ij}\to0$.

\subsection{The Yield Surfaces\label{yield surfaces}} 

In a space spanned by stress components and the density, there is a surface  that 
divides two regions in any granular media, one in which the grains necessarily move, 
another in which they may be at rest. We shall  refer to this surface as {\em the 
yield surface} -- though we emphasize that it is unrelated to, and different from, any yield associated with the critical state, see the next paragraph.  To make its definition  precise, we take the yield surface 
to be the divide between two regions, one in which elastic solutions may be stable, and another in which they never are. Clearly, the medium may be at rest for a given stress only if an appropriate elastic solution is stable. 
Since the elastic energy of any solution satisfying Eq~(\ref{2a-1}) is extremal, the energy is convex and minimal in the stable region, concave and maximal in the unstable one ---in which infinitesimal perturbations suffice to destroy the solution.

The yield surface defines a yield stress [such as given by Eq~(\ref{2b-3}) below]. Many textbooks identify it with the highest shear stress achieved in an approach to the {\em critical state}, with the justification that the accompanying  shear rate is so low that one may consider the motion {\em quasi-static}. And since the critical state is a form of yield, the physics behind it must be static, energetic. We believe this argument overlooks the following point: 
A  {quasi-static motion} is one that visits a series of static, equilibrium states, so slowly that the dissipation is negligible, implying $T_g\to0$. This is what was defined as {\em quasi-elastic motion} above, see also Sec~\ref{3regimes}. The rate-independent,  hypoplastic motion, taking place  eg. during an approach to the critical state, is different. It does visit a series of elastic states, but at an elevated $T_g$, and is therefore highly dissipative.
The energetic instability and the critical state are two distinct concepts, static versus dynamic. 
The first is a convexity transition of the elastic energy, the second a stationary solution of  the 
evolution equation for the elastic strain $u_{ij}$, see Sec~\ref{uri-regime}, comparable to the stationary solution of any diffusion equation. 
The two yield stresses are frequently similar in magnitude, which is probably  related
to the fact that both account for the clearance with the
profile of the underlying layer, though one with granular jiggling, $T_g\not= 0$, and hence a little easier. But the yield stress given by Eq~(\ref{2b-3}) below needs to be larger than the highest shear stress achieved during an approach to the critical state, because  a series of elastic states is being visited during  the approach. Otherwise, the system will abandon it, in search for a stable but nonuniform configuration, typically shear bands.

\subsubsection{The Coulomb Yield Surface\label{Druck-Prager}}
The elastic energy of Eq~(\ref{2b-2}) is convex only for 
\begin{equation}\label{2b-3} u_s/\Delta\le\sqrt{2{\cal B}/{\cal A}} \quad
\text{or}\quad \pi_s/P_\Delta\le\sqrt{2{\cal A}/{\cal B}},
\end{equation} 
turning concave if the condition is violated. The second constraint may be derived by rewriting Eq~(\ref{2b-2b}) as 
${4P_\Delta}/{\pi_s}={2{\cal B}{\Delta}}/{{\cal
A}}{u_s}+{u_s}/{\Delta}$, 
which shows
$P_\Delta/\pi_s=\sqrt{{\cal B}/2{\cal A}}$ is minimal for
$u_s/\Delta=\sqrt{2{\cal B}/{\cal A}}$. This corroborates the behavior that no
granular system stays static if the shear stress is too large for given
pressure. We call it the Coulomb yield surface, although technically, it is the Drucker-Prager relation, see Sec.\ref{MYS}. And again, nothing in connection to the critical state is  meant here.

Taking ${\cal B}/{\cal A}$ as density independent, typically 
${\cal B}/{\cal A}\approx5/3$, 
we only need to specify the density dependence of $\cal B(\rho)$, which we require should account for the following three important characteristics of granular media:
\begin{itemize}
\item The energy should be concave for $\rho<\rho_{\ell p}$, the
random loose density, as no elastic solution exists when the grains  loose contacts with one another.
\item The energy must be convex for larger densities, $\rho_{\ell p}<\rho<\rho_{cp}$, to ensure the stability of elastic solutions in this region. 
\item The density dependence of sound
velocities as measured by Harding and Richart~\cite{hardin} should be well 
rendered by  $\sqrt{\partial^2w/\partial u_s^2}\sim\sqrt{\cal B}$. 
\end{itemize}
The simplest expression we could find [see~\cite{granR2} for details of the struggle] is  
\begin{eqnarray}\label{2b-4} {\cal B}={\cal B}_0
[(\rho-\bar\rho)/(\rho_{cp}-\rho)]^{0.15},\quad
\bar\rho\equiv(20\rho_{\ell p}-11\rho_{cp})/9, 
\end{eqnarray} 
with ${\cal B}_0>0$ a material constant. 
The small exponent of 0.15 does not imply an accuracy over a few orders of magnitude for $\rho\to\bar\rho$. Since $\cal B$ loses its convexity at $\rho_{\ell p}$, the density is never close to $\bar\rho$. (Note $\bar\rho<\rho_{\ell p}<\rho_{cp}$, with  $\rho_{cp}-\rho_{\ell p}\approx\rho_{\ell p} -\bar\rho$.) And although $\rho$ may be close to $\rho_{cp}$, the slow divergence only expresses, qualitatively and very tentatively, that the system becomes orders of magnitude stiffer there. 

\subsubsection{More and Different Yield Surfaces\label{MYS}}
\begin{figure}[t] \begin{center}
\includegraphics[scale=0.28]{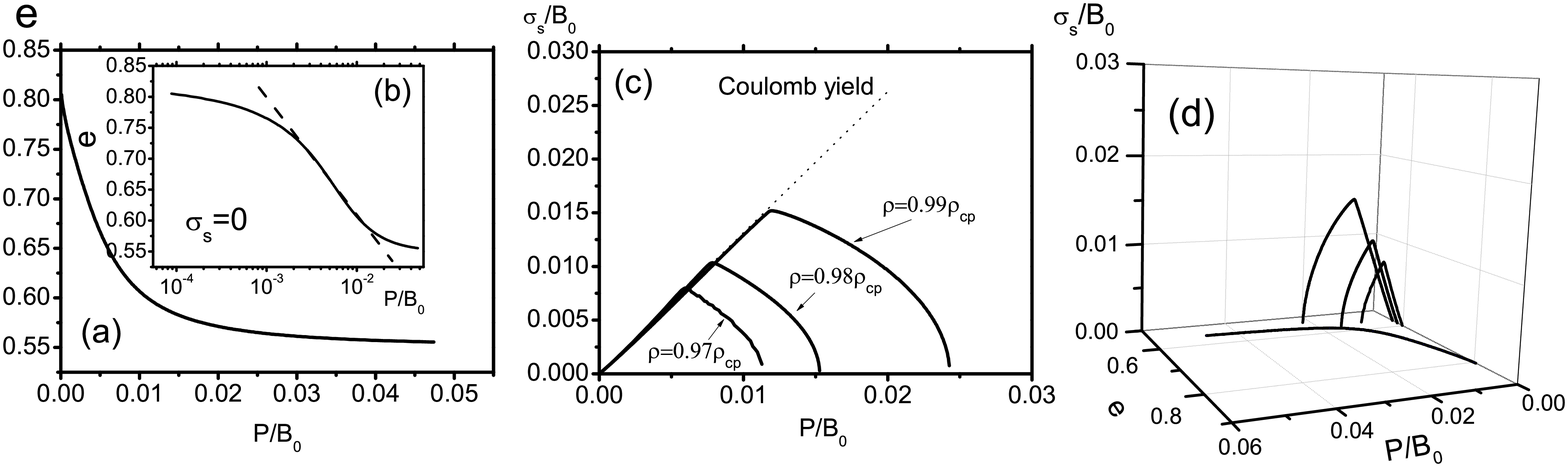}
\end{center}
\caption{\label{fig1}
Granular yield surfaces for $T_g=0$, as a function of the pressure $P$, shear stress $\sigma_s=\pi_s$, and the void ratio $e$, as calculated from the convexity transition of the energy given in Eqs.(\ref{2b-2},\ref{2b-4},\ref{2b-5a}), with ${\cal B}/{\cal A} =5/3$, ${\cal D}_{1}=1$, ${\cal D}_{2},{\cal D}_3=2$, and $\rho _{lp}=0.85\rho _{cp}$.
The plots (a,b) are at $\sigma _{s}=0$, with the inset having a logarithmic scale; the curves of (c) are at the indicated densities. [The dashed straight lines in (b,c) are, respectively, the formula $e=e_{0}-k\ln P$ and the Coulomb yield line.] The curves of (d) are the same as in (a,c), though now in 3D-space, spanned by $e,P,\sigma_s$. }
\end{figure} 

As depicted in Fig.~\ref{fig1},  granular media possess more yield surfaces. We consider
the space spanned by the pressure $P_\Delta$, shear stress $\sigma_s=\pi_s$, and the void ratio $e$, where $e\equiv1/\phi-1$. ($\phi\equiv\rho/\rho_g$ is the packing fraction, and $\rho_g$  the bulk density of the grains.) First,
for given $e$, there should be a maximal pressure that a granular system can
sustain before it collapses, implying a yield surface as depicted in (a) of Fig~\ref{fig1}. 
Sand at rest will not cross this  boundary when compressed. Instead, it will
collapse, becoming more compact, with a smaller $e$. The curve $e(P)$ in (a) holds for vanishing shear stress $\sigma_s=0$. If there were no dependence of  $\sigma_s$, we would have vertical lines in (c), connecting the $P$-axis and the Coulomb yield line, the position of which depends on $e$. More plausible, however, would be a bending of these lines, as depicted, because a shear stress should render a static granular ensemble less stable.   All this may be accounted for in {\sc gsh} by higher order terms in the elastic energy. 

Although the qualitative aspects of the above described behavior must be correct, it is difficult to make them more quantitative. For lack of better data, we tentatively identify the behavior of (a) with what in textbooks on  soil-mechanics~\cite{schofield,nedderman} is frequently referred to as the {\em virgin consolidation line}, and that of (c) with ``{\it caps}.'' This may not be appropriate, because both are usually associated with clay, and there are indications that with sand the consolidation line is associated with grain crushing~\cite{einaf}.  [The inset, (b) of Fig~\ref{fig1}, has a logarithmic scale. It serves to demonstrate that the standard formula  $e=e_{0}-k\ln P$ do not go to $\rho _{lp}$ and $\rho _{cp}$,  for $P\rightarrow 0$ and $\infty $, respectively.] 

We include the following higher- order terms, with ${\cal D}_1,{\cal D}_2, {\cal D}_3>0$,
\begin{equation}\label{2b-5a} -{\cal B}_0({\cal D}_1\Delta^3
+{\cal D}_2\Delta u_s^2+{\cal D}_3 u_s^4), 
\end{equation} 
to be added to $w_\Delta$,
Eq~(\ref{2b-2}). Consider first $u_s^2=0$. If $\Delta$ is large enough, the
term $-{\cal D}_1\Delta^3$, with a negative second derivative, will
work against ${\cal B}\Delta^{2.5}$ and turn $w_\Delta$ concave. The value 
$\Delta_c$ at which this happens is given by
$\sqrt{\Delta_c}=5{\cal B}(\rho)/8{\cal D}_1(\rho)$. As ${\cal B}$ diverges
at $\rho_{cp}$, so does $\Delta_c$. If $\Delta_c(\rho)=0$ for
$\rho=\rho_{\ell p}$, ${\cal D}_1(\rho)$ will have to diverge there.
Next consider $u_s^2\not=0$. If ${\cal D}_2,{\cal D}_3=0$, the yield lines in the
space spanned by $P_ \Delta,\pi_s$ for given density would be vertical
lines. The presence of $-{\cal D}_2\Delta u_s^2$ and  $-{\cal D}_3 u_s^4$ reduce the value of
$\Delta$ (or $P_ \Delta$) for growing $u_s$ (or $\pi_s$), bending the lines
to the left. We did not find enough data that we could have used to fix the values of ${\cal D}_1,{\cal D}_2,{\cal D}_3$.

Next we address varying forms of yield laws, of which there are many. That of Eq~(\ref{2b-3}) is usually referred to as the
Drucker-Prager approximation of the Coulomb yield surface. The actual Coulomb 
law is anisotropic. And there are those referred to as 
{\em Lade-Duncan}~\cite{lade-duncan} or {\em Matsuoka-Nakai}~\cite{matsuoka}.
Defining the friction angle as
$\varphi\equiv\arcsin\sqrt{3/(6P_\Delta^2/\pi_s^2-1)}$, the Coulomb,
Drucker-Prager, and Lade-Duncan yield laws are respectively given as
\begin{eqnarray}\nonumber  \frac{\pi_3-\pi_1}{\pi_3+\pi_1}
=\sin\varphi,
\quad
\frac{\pi_s}{P_\Delta}=\frac{\sqrt{6}\sin\varphi}{\sqrt{3+\sin^2\varphi}},
\quad\frac{\pi_1\pi_2\pi_3}{27P^3_\Delta}
=\frac{(1-\sin\varphi)\cos^2\varphi}{(3-\sin\varphi)^3}. \end{eqnarray} 
Engineers choose among them depending on the system, personal preferences
and experiences, apparently without a commonly accepted rule. 
We discovered that, by including
the third strain invariant $u_t^3\equiv u^*_{ij}u^*_{jk}u^*_{ki}$ into
Eq~(\ref{2b-2}), 
\begin{equation}\label{2b-6} w_\Delta= {\mathcal B}
\sqrt{\Delta }\left(2\Delta^2/5+ {\mathcal A} u_s^2/ {\mathcal B}
-{\mathcal C} {u^3_t}/ {\mathcal B}\Delta\right), \end{equation}
with ${\cal A,B,C}>0$, it is possible to account for all these laws
simultaneously. (Note the new term is also of order 2.5.) Tuning $\cal C$ is, the yield surface an be made numerically
indistinguishable from all these yield laws. Because a single expression
is employed, and because intermediate yield laws are also possible, this is
a simplifying and unifying step, see~\cite{3inv} for details (including how $u_{ij},\pi_{ij}$ remain collinear). 

\subsection{Dynamics \label{dynamics}} 
\subsubsection{Structure of the Dynamics}

Next, we specify the evolution equations for the state variables. The equation for the elastic strain, assuming both $u_{ij},v_{ij}$ are uniform, is~\cite{granR2}
\begin{equation}\label{2c-6a}
\partial_tu_{ij}-v_{ij}+\alpha_{ijk\ell}v_{k\ell}=-(\lambda_{ijk\ell}T_g)\,u_{k\ell}
\end{equation} 
(where  $v_{ij}\equiv\frac12(\nabla_iv_j+\nabla_jv_i)$ is the shear rate,
$v^*_{ij}$ its traceless part, and $v_s^2\equiv v^*_{ij}v^*_{ij}$).  
If $T_g$ is finite, grains jiggle and briefly
lose contact with one another, during which their deformation is
partially lost. (More realistically, the grains only loosen contact with one another. But this suffices to mobilize them, and free them briefly from the ``elastic corset'' of neighboring grains that maintains the deformation.) Macroscopically, this shows up as a relaxation of
$u_{ij}$, with a rate that grows with $T_g$, and vanishes for
$T_g=0$. So the lowest order term in a $T_g$-expansion is
$\lambda_{ijk\ell}T_g$. With the elastic energy a convex function, the (negative) elastic stress $-\pi_{ij}=\partial w/\partial u_{ij}$ is a monotonically increasing function of $u_{ij}$. Therefore,  $-\pi_{ij}, u_{ij}$
decrease at the same time. And Eq~(\ref{2c-6a}) accounts for the stress relaxation discussed in the introduction. 

The Onsager coefficient $\alpha_{ijk\ell}$ is an off-diagonal element. Dividing $u_{ij}$ into $\Delta\equiv-u_{\ell\ell}$, $u_{ij}^*$, and specifying the matrices $\alpha_{ijk\ell},\lambda_{ijk\ell}$ with two elements each, Eq~(\ref{2c-6a}) is written as
\begin{eqnarray}
\label{2c-7xx}
\partial_t\Delta+(1-\alpha )v_{\ell\ell} -\alpha_1u^*_{ij}v^*_{ij}
=-\lambda_1T_g\Delta, 
\\\label{2c-8} 
\partial_tu^*_{ij}-(1-\alpha )v^*_{ij}
= -\lambda T_gu^*_{ij},
\\\label{2c-9xx}
\partial_tu_s-(1-\alpha )v_s= -\lambda T_gu_s.
\end{eqnarray} 
The third equation is valid only if strain and rate are collinear, $u^*_{ij}/|u_s|=v^*_{ij}/|v_s|$. This is frequently the case for steady rates, because any 
component of $u_{ij}$ not collinear with $v_{ij}$ relaxes to zero. 
The coefficient  $\alpha$ (assuming  $0<\alpha <1$) describes a  reduced gear ratio: The same shear rate yields a smaller deformation, $\partial_tu_{ij}=(1-\alpha)v_{ij}\cdots$, but 
acts also at a smaller stress, $\sigma_{ij}=(1-\alpha)\pi_{ij}\cdots$, see below.  $\alpha_1$ accounts for the fact that shearing granular media will change the compression $\Delta$, implying {\em dilatancy} and {\em contractancy}. (Although more Onsager coefficients are permitted by symmetry, they have been excluded to keep the equations as simple as possible.) 

Next are the continuity equations for mass and momentum
density, 
\begin{equation}\label{2c-1} \partial_t\rho+\nabla_i(\rho
v_i)=0,\qquad \partial_t(\rho v_i)+\nabla_j(\sigma_{ij} +\rho v_iv_j)=0,
\end{equation} 
where the stress tensor $\sigma_{ij}=P\delta_{ij}+\sigma^*_{ij}$ (with $\sigma^*_{ij}$ the traceless part) is determined by general principles~\cite{granR2,granR3}  as
\begin{eqnarray} \label{2c-2}
P\equiv\sigma_{\ell\ell}/3=(1-\alpha )P_\Delta+P_T-\zeta_gv_{\ell\ell},
\\\label{2c-2a}
\sigma^*_{ij}=(1-\alpha)\pi_{ij}^*-\alpha_1u^*_{ij}P_\Delta -\eta_gv^*_{ij},
\\ \label{2c-2b}
\sigma_{s}=(1-\alpha )\pi_s-\alpha_1u_sP_\Delta
-\eta_gv_s. 
\end{eqnarray} 
Again, the third equation (with $\sigma_s^2\equiv\sigma_{ij}^*\sigma_{ij}^*$) is
valid only if $\pi_{ij}^*$ and $v^*_{ij}$ are collinear,
$\pi_{ij}^*/|\pi_s|=v^*_{ij}/|v_s|$. 
The pressure $P$ and shear stress $\sigma_s$ contain elastic contributions
$\sim\pi_s,P_\Delta$ from Eq~(\ref{2b-2b}), the seismic pressure $P_T\sim T_g^2$ from Eq~(\ref{2b-5}), and viscous contributions $\sim\eta_g,\zeta_g$. The  off-diagonal Onsager coefficients $\alpha ,\alpha_1$ (introduced in the equation for the elastic strain $u_{ij}$) soften and mix the elastic stress components. The term preceded by $\alpha_1$ is smaller by an order in the elastic strain, and may frequently be neglected.

The balance equation for the granular entropy $s_g=b\rho T_g$ is 
\begin{eqnarray}\label{2c-4} 
\partial_ts_g+\nabla_i(s_gv_i-\kappa\nabla_iT_g)=
(\eta_g v_s^2+\zeta_gv^2_{\ell\ell}-\gamma T_g^2)/T_g. 
\end{eqnarray}
Here, $s_gv_i$ is the convective, and $-\kappa\nabla_iT_g$ the diffusive flux. $\eta_g v_s^2$ accounts for viscous heating, for the increase of $T_g$ because 
macroscopic shear rates jiggle the grains. A compressional rate
$\zeta_gv^2_{\ell\ell}$ does the same, though not as efficiently~\cite{granL3}. The term $-\gamma T_g^2$ accounts for the relaxation of $T_g$, ie., for the conversion of granular energy into inner granular one. Frequently, this equation may be simplified, first 
by linearizing in $\nabla_iT_g$, assuming it to be small; then by taking all other variables to be uniform,  the  convective term and $v_{\ell\ell}$ as negligible. Finally, an extra source term $\gamma_1 h^2T_a^2$ may be added, to account for an ``ambient temperature'' $T_a$ -- external perturbations such as given by a sound field or by tapping. 
(Generally speaking, any source mechanism contributing to $T_g$ is already included in the expression without $T_a$. For instance, given a sound field -- generated either by loudspeakers or tapping -- there is the term on the right hand side 
of Eq~(\ref{2c-4} ),  $\zeta_1 (v_{\ell\ell}^{sound})^2$, where $v_{\ell\ell}^{sound}$ is the fast varying compressional rate of the sound field. Coarse-graining it, we may set 
$\langle\zeta_1 (v_{\ell\ell}^{sound})^2\rangle \equiv\gamma_1h^2T_a^2\equiv \eta_1 v_a^2$, 
to quantify this contribution, either in terms of $T_a$, or the shear rate $v_a$ needed to produce this $T_a$. Adding such a term is a convenient short cut to account for a general perturbation without specifying the cause.)  The result is
\begin{eqnarray}
\label{sum T_g} 
b\rho\partial_tT_g-\kappa_1T_g\nabla^2T_g=
\eta_1 v_s^2-\gamma_1 h^2(T_g^2-T_a^2). 
\end{eqnarray}
An rather similar equation holds for the true entropy $s$, see~\cite{granR2}.

\subsubsection{Transport Coefficients\label{transport coefficients}}
All coefficients $\alpha,\alpha_1,\eta_g,\zeta_g$ are functions of the state variables, $u_{ij}$, $T_g$ and $\rho$. As the hydrodynamic formalism only delivers 
the structure of the dynamics, not the functional 
dependence of the  transport coefficients, these are to be obtained (same as the energy) from experiments or simulations, in an iterative  trial-and-error process. And the specification below is what we at present believe is appropriate. Generally speaking, we find 
strain dependence to be weak -- plausibly so because the strain is a small
quantity. We expand in it, keeping only the constant terms. We also expand in $T_g$, but eliminate the constant terms, because we assume granular media are fully elastic for $T_g\to0$, implying the force balance $\nabla_j\sigma_{ij}=\rho{\rm g}_i$ should reduce to the equilibrium condition, Eq~(\ref{2a-1}). Therefore we take
$\alpha,\alpha_1,\eta_g,\zeta_g,\kappa_g$ to vanish for $T_g\to0$. In addition, we
also need $\alpha,\alpha_1$ to saturate at an elevated $T_g$, such that rate-independence may be established in the hypoplastic regime. Hence
\begin{eqnarray}\label{2c-3}
\eta_g=\eta_1T_g,\,\, \zeta_g=\zeta_1T_g,\,\,
\kappa=\kappa_1T_g,
\\\nonumber
\alpha/\bar\alpha =
\alpha_1/\bar\alpha_1={T_g}/({T_\alpha+T_g}),
\end{eqnarray} 
with $\bar\alpha,\bar\alpha_1,\eta_1,\zeta_1,\kappa_1,T_\alpha$ functions of $\rho$ only, or  the packing fraction $\phi$. Expanding $\gamma$ in $T_g$, 
\begin{equation}\label{2c-5} \gamma=\gamma_0+\gamma_1 T_g, 
\end{equation} 
we keep $\gamma_0$, because the reason that led to Eqs~(\ref{2c-3}) does not apply. More importantly, $\gamma_0$ ensures a smooth transition from the hypoplastic to the quasi-elastic regime, see Eq~(\ref{TgVs2}) below. (Although $\gamma_0=0$ in rarefied systems~\cite{luding2009}, we do not see any reasons for this to hold for denser ones.) 

The transport coefficients are also functions of $\rho$, containing especially a divergent/vanishing  part $\sim(\rho_{cp}-\rho)$. Assuming that, at $\rho=\rho_{cp}$, the plastic phenomena of stress relaxation, softening and dilatancy vanish, $T_g$ relaxes instantly, and the system is infinitely viscous, we take 
\begin{equation}\label{density-dependence}
\lambda,\,\lambda_1,\,\alpha,\,\alpha_1,\,\gamma_1^{-1},\,\eta_1^{-1}\sim\rho_{cp}-\rho.
\end{equation}  
We need to stress that we stand behind the temperature dependence with much more confidence than that of the density, for two reasons: First, $\rho$ is not a small quantity that one may expand in, and we lack the general arguments employed to extract the $T_g$-dependence. Second, not coincidentally, the $\rho$ dependence does not appear universal: The above dependence of $\gamma_1,\,\eta_1$ seems to fit glass beads  data, while $ \gamma_1\sim(\rho_{cp}-\rho)^{-0.5}$, $\eta_1\sim(\rho_{cp}-\rho)^{-1.5}$ appear more suitable for polystyrene beads, see~\cite{denseFlow}.

At given shear rates, $v_s=$ const, the stationary state of Eq~(\ref{2c-4}) -- characterized by $\partial_ts_g=0$,  with viscous heating  balancing $T_g$-relaxation -- is quickly
arrived at, say within $10^{-3}$ s in dense granular media, implying 
\begin{eqnarray}\label{2c-6}
{\gamma_1} \,h^2\, T_g^2=v_s^2\,{\eta_1}+v^2_{\ell\ell}\,{\zeta_1},
\quad
h^2\equiv1+\gamma_o/(\gamma_1T_g).
\end{eqnarray} 
Taking the density for simplicity as either constant or slowly changing, ie. $v^2_{\ell\ell}\approx0$, 
we have a quadratic regime for small $T_g$ and low $v_s$, and a linear one at elevated $T_g,v_s$:
\begin{eqnarray}
\label{TgVs} 
T_g=|v_s|\sqrt{\eta_1/\gamma_1}\quad\,\,\text{for}\quad \gamma_1T_g\gg\gamma_0,
\\\label{TgVs2} 
T_g=v_s^2\,\,({\eta_1/\gamma_0})\quad\text{for}\quad \gamma_1T_g\ll\gamma_0.
\end{eqnarray}
As mentioned above and discussed in the next section, the linear regime is hypoplastic, in which the system displays rate-independent elasto-plastic behavior and the hypoplastic model holds. In the quadratic regime, because $T_g\sim v_s^2\approx0$ is quadratically small,  the behavior is quasi-elastic, quasi-static, with slow, consecutive visit of static stress distributions. Note that we have $h=1$ in the hypoplastic regime, and $h\to\infty$ in the quasi-static one.   

Eqs~(\ref{2c-7xx},\ref{2c-8},\ref{2c-9xx}) also have a stationary solution, 
$\partial_t\Delta, \partial_tu_s=0$, in which the shear rate $v_s=$ const 
is compensated by the relaxation $\sim T_g$. As a result, 
$\Delta=\Delta_c,u_s=u_c$ remain constant, and with them also the 
pressure and shear stress, $P=P_c, \sigma_s=\sigma_c$. This ideally plastic behavior is the {\em critical state}. In the linear regime,
$T_g\sim|v_s|$, both $P_c$ and $\sigma_c$ are rate-independent. Since the rate-independent critical state is a motion in the linear regime, and since it is irreversible and strongly dissipative, it is not quasi-static.

\subsection{Summary\label{sum}}

With the above set of equations derived, the expressions for energy density and transport coefficients in large part specified, {\sc gsh} is a fairly well-defined theory. It contains clear ramifications and provides little 
leeway for retrospective adaptation to observations. As a first step to coming to terms with its  ramifications, we examine its basic features. 

Granular rheology as observed may be divided into three shear rate regimes: {\em Bagnold} for high, {\em hypoplastic} for low, and {\em quasi-elastic} for even lower 
ones. Fast dense flow is in the first regime, in which pressure and shear stress are proportional to shear rate squared, $p,\sigma_s\sim v_s^2$. Various elasto-plastic motions, observed especially 
in triaxial apparatuses, are in the second, rate-independent regime. The third regime is elastic -- no difference between load and unload, and no critical state. 
Static stress distribution and elastic waves belong here. 
This third regime is again rate-independent. 
Although textbooks, taking the hypoplastic regime as quasi-static, do not acknowledge the existence of a third rate regime, we note that elastoplastic motion cannot be quasi-static, because it is plastic and irreversible, see the discussion in Sec~\ref{yield surfaces} and~\ref{3regimes}.  In {\sc gsh}, the static, equilibrium state with $T_g=0$ is fully elastic.  If quasi-static motion exists, it must be quasi-elastic. On the other hand, it is admittedly difficult to observe. Some possible reasons are discussed in Sec~\ref{aeipt}, with suggestions in~\ref{soft springs} on how to overcome them. 

{\sc gsh} is constructed such that any deviation from elasticity -- encapsulated in the 
coefficients $\alpha,\alpha_1,\eta_g,\zeta_g,\kappa_g,\lambda
T_g,\lambda_1T_g$ -- vanishes with $T_g$.
For $T_g=0$, we have $\partial_t u_{ij}=v_{ij}\equiv\partial_t\epsilon_{ij}$, or
$u_{ij}=\epsilon_{ij}$,
$\sigma_{ij}=\pi_{ij}$, 
implying perfect elasticity. At very low shear rates, $T_g\sim v_s^2$, deviations from elasticity are quadratically small. The system is then {quasi-elastic} -- though only as long as no yield surface (as discussed in Sec~\ref{yield surfaces}) is breached. 

When $T_g$ is more elevated, we are in the linear regime, $T_g\sim|v_s|$, see Eq~(\ref{TgVs}). Here, the full complexity of granular media emerges. Nevertheless, three scalar equations, derived starting from {\em two-stage irreversibility} and {\em variable transient elasticity}, suffice to account for most phenomena. Two account for transient elasticity, 
Eqs~(\ref{2c-7xx},\ref{2c-9xx}), and one for $T_g$, Eq~(\ref{2c-4}) or Eq~(\ref{sum T_g})
In the hypoplastic regime, the stress is still elastic, though softened by  $\bar\alpha$. 
Noting $\pi^*_{ij}$, $u^*_{ij}, \sigma^*_{ij}$ are
collinear, and assuming the higher order term $\bar\alpha_1u_sP_\Delta$ may be neglected, we have the rate-independent expressions
\begin{equation}
P=(1-\bar\alpha)P_\Delta, \quad \sigma_{s}=(1-\bar\alpha)\pi_s.
\label{sum stress hp}\end{equation}
As we shall see, these simple expressions are well capable of accounting for elasto-plastic motion generally, including especially load-unload behavior, Sec~\ref{Load and Unload}, and the approach to the critical state, Sec~\ref{critical state}. They were also used for a successful comparison to the hypoplastic and barodesy model, in Sec~\ref{conrel}, and for the damping of elastic waves, Sec~\ref{elastic waves}.

For yet larger rates, the total stress includes the seismic pressure $P_T$ and the viscousity (of which the compressional one is neglected), see Eqs~(\ref{2c-2},\ref{2c-2b}),
\begin{eqnarray}\label{sum stress df}
P=(1-\bar\alpha)P_\Delta+{\textstyle\frac12} T_g^2\, a\, \rho^2 \,b/(\rho_{cp}-\rho),
\quad
\sigma_{s}=(1-\bar\alpha)\pi_s-\eta_1T_gv_s.
\end{eqnarray}
Since $T_g\sim|v_s|$,  we have $P_T\sim T_g^2\sim v_s^2$ and $\eta_gv_s=\eta_1T_gv_s\sim v_s^2$. 
So both may be written as $e_1+e_2v_s^2$, implying a quadratic
dependence on the rate for $e_2v_s^2\gg e_1$, and rate-independence for  $e_2v_s^2\ll e_1$. Rapid dense flow is considered in~\cite{p&g2013}.
This ends the brief presentation of {\sc gsh}.


\section{The Quasi-Elastic Regime\label{quasi elastic motion}}

\subsection{Quasi-Elastic versus Hypoplastic Regime\label{3regimes}} 

Many in soil mechanics call the slow granular motion in the hypoplastic  regime -- say the approach to the critical state -- {\it quasi-static}. We do not think this is the right term, because, as discussed at the beginning of Sec.\ref{yield surfaces}, the motion occurs at an elevated $T_g$, is dissipative and irreversible. 
Quasi-static motion is never dissipative. Consider sound propagation in any system, say Newtonian liquid, elastic media or liquid crystals. The velocity is a constant, and the damping $\sim\omega$, the frequency. Therefore, sound waves are less damped the smaller the frequency is. This is a rather generic feature: Changing the state variable $A$ slowly, dissipation vanishes with $\partial_tA$, the rate of change. At the very slow limit, the dissipation may be neglected, and the motion is rate-independent. It is then called  {\em quasi-static}, because the system is at this rate visiting static, equilibrium states consecutively.  

Granular systems are both dissipative and rate-independent in the hypoplastic regime. As we shall see in Sec~\ref{uri-regime}, this rate-independence is a reflection of the fact that  reactive and dissipative terms have the same frequency dependence, and are comparable in size -- they are exactly equal in the critical state. If there were only the hypoplastic regime, elastic waves would always be overdamped. Since this is not the case, there must be a different rate-independent but dissipation-free regime.  
Faced with this dilemma, a frequent suggestion is to take a small incremental strain (such as given in an elastic wave) to be elastic and free of dissipation, but a large one as elasto-plastic and dissipative. For the following reason, we believe this is incompatible with the notion of a quasi-static motion, and  the wrong way out: Starting from a static state of given stress, and applying a small incremental strain that is elastic, the system is again in a static state and an equally valid starting point. The next small increment must therefore also be purely elastic. Many consecutive small increments yield a large change in strain, and if the small ones are not dissipative, neither can their sum be. 

In {\sc gsh}, it is the strain rate, not its amplitude, that decides whether the system is elastic or hypoplastic.  Small strain increments achieved with
a high but short lasting shear rate will provoke an elastic response, if
$T_g$ does not have time to get to a sufficiently high value to induce
any plastic responses. Furthermore, the mere existence of a quasi-static, quasi-elastic regime does not imply that it is also easily observable, though see Sec~\ref{soft springs}.
%

Finally, we note that backtracing of the stress curve $\hat\sigma(t)$ when reversing the strain, $\hat\epsilon(t)\to\hat\epsilon(-t)$, occurs only in
the quasi-elastic regime, not the hypoplastic one. (We use a hat
to indicate a tensor.)  The stress is a function of the
elastic strain, $\hat\sigma=\hat\sigma(\hat u)$. Reversing $\hat u(t)$ will
always backtrace $\hat\sigma(t)$. But only in the quasi-elastic regime may
we identify $\hat u(t)=\hat\epsilon(t)$. Failure to backtrace at hypoplastic rates are not evidence of a ``history dependence."

\subsection{An Elastic-Ideally-Plastic Trajectory\label{aeipt}}
In the quadratic regime of very slow shear rates,  $T_g\sim |v_s|^2\to0$, the granular temperature is so small that the system is essential elastic, moving from one elastic, equilibrium state to a slightly different one. This is the reason we call it {\em quasi-elastic}. Because $\hat\sigma=\hat\pi$ and $\partial_t\hat u\to\partial_t\hat\epsilon=\hat v$, the change of the the shear stress $\sigma_s$ is well approximated by the (hyper-) elastic relation, 
\begin{equation}\label{3a-1}
\partial_t\sigma_{ij}=\frac{\partial\sigma_{ij}}{\partial
u_{k\ell}}\partial_t u_{k\ell} =\frac{\partial\pi_{ij}}{\partial
u_{k\ell}}\partial_t\epsilon_{k\ell}=-\frac{\partial^2 w}{\partial
u_{ij}\partial u_{k\ell}}v_{k\ell}. 
\end{equation} 
Shearing a granular medium at quasi-elastic rates, the result will be a trajectory $\hat\sigma(\hat\epsilon)$ that is much steeper than in experiments at hypoplastic rates, such as observed during an approaching to the critical state. The gradient is given directly by the stiffness constant ${\partial^2 w}/{\partial \hat u^2}$, and possibly three to four times as large as the average between loading and unloading at hypoplastic rates [because Eq~(\ref{2c-8}) lacks the factor of $(1-\alpha)$]. This goes on until the system reaches a yield surface of the elastic energy, the convexity transitions  discussed in Sec~\ref{yield surfaces}. The system becomes ideally plastic at this point, abruptly, by forming shear bands. The critical state will not be reached. Reversing the shear rate in between will retrace the function $\hat\sigma(t)$. 

\subsection{Soft Springs versus Step Motors\label{soft springs}}
\begin{figure}[b] \begin{center}
\includegraphics[scale=0.35]{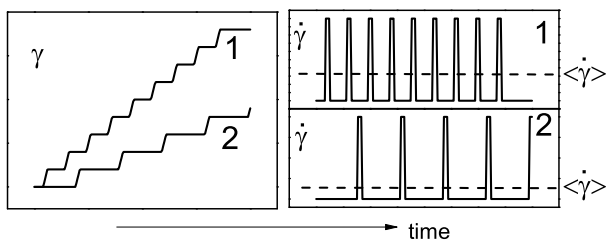}
\end{center}
\caption{Why observing the quasi-elastic regime is hard if step motors are used.\label{StepMotor}} 
\end{figure} 
Quasi-elastic behavior has not been observed in triaxial apparatus. This may simply be because  even the lowest rates are not slow enough. Or because step motors are widely used in these appliances. Plotting the shear 
rate versus time, $\dot\gamma(t)$, different shear rates are approximately given as depicted in Fig~\ref{StepMotor}. Although the curves have different average rates $\langle\dot\gamma\rangle$, the time-resolved, maximal rates  $\dot\gamma_M$ are identical. And if the time span of  $\dot\gamma_M$ is long enough for $T_g$ to respond, and $\dot\gamma_M$ is high enough for the system to be in the linear regime, $T_g\sim \dot\gamma_M$, the system will display consecutive hypoplastic behavior in both cases, irrespective of  the average rate $\langle\dot\gamma\rangle$. 

Here, we suggest two ways to observe quasi-elastic behavior, both by fixing the stress rate at low $T_g$, because a given stress rate corresponds to two different shear rates, a high one at elevated $T_g$ and a low one at vanishing $T_g$.  The first method is slowly incline a plane supporting a layer of grains. In such a situation, the shear rate remains very small, and the system starts flowing only when a yield surface is breached. In contrast, employing a feedback loop in a triaxial apparatus to maintain a stress rate would not work well, because the correcting motion typically has strain rates that are too high.

A second method is to insert a very {soft spring}, even a rubber band, between the granular
medium and the device moving at a given velocity $v$ to deform it. If the spring is softer by a large factor $a$ than the granular medium (which is itself rather soft), it will absorb most of the displacement, leaving the granular medium deforming at a rate smaller by
the same factor $a$ than without the spring. In other words, the soft spring serves as a ``stress reservoir'' for the granular medium. The same physics applies when the feedback loop is connected via a soft spring, as then only little $T_g$ is excited.

\section{The hypoplastic regime\label{uri-regime}}

Hypoplastic motion occurs at an elevated $T_g\sim|v_s|$, in what we have named the linear regime. It is  {\em rate-independent} for given, constant strain rates, in the sense that the increase in the stress $\Delta\sigma_{ij}$ depends only on the increase in the strain,
$\Delta\epsilon_{ij}=\int v_{ij}{\rm d}t$, not how fast it takes place. We also call this regime hypoplastic because this is where the {\em hypoplastic model} holds, a state-of-the-art engineering theory~\cite{kolymbas1} that we shall consider in Sec~\ref{Hypoplasticity}.

\subsection{Load and Unload\label{Load and Unload}} 

In the hypoplastic regime, for given shear rate $v_s$, the granular temperature relaxes quickly to its stationary value  $T_g=|v_s|\,\sqrt{\eta_1/\gamma_1}$. Inserting this into
Eqs~(\ref{2c-7xx}, \ref{2c-9xx}), we arrive at
\begin{eqnarray}
\partial_t\Delta=v_s\,\alpha_1u_s&-&|v_s|\,\Lambda_1\Delta, 
\quad \label{3b-3}
\partial_tu_s=v_s\,(1-\alpha )-|v_s|\,\Lambda u_s, 
\\\label{3b-4}
\Lambda&\equiv&{\lambda}/{h}\sqrt{{\eta_1}/{\gamma_1}}\equiv\Lambda_1{\lambda}/{\lambda_1}\sim
(\rho_{cp}-\rho), \end{eqnarray} 
which are explicitly rate-independent for $\alpha=\bar\alpha,\alpha_1=\bar\alpha_1$, see Eq~(\ref{2c-3}). 
The last equation is a result of inserting the density dependence of Eqs~(\ref{density-dependence}) and indicates that relaxation of the elastic strain becomes slower at higher density, and stops at the close-packed density $\rho_{cp}$, where the system is essentially elastic. We take  $\Lambda\approx3.3\Lambda_1$, as compressional relaxation is typically slower than shear relaxation~\cite{granL3}. 

\begin{figure}[tbh] \begin{center}
\includegraphics[scale=0.3]{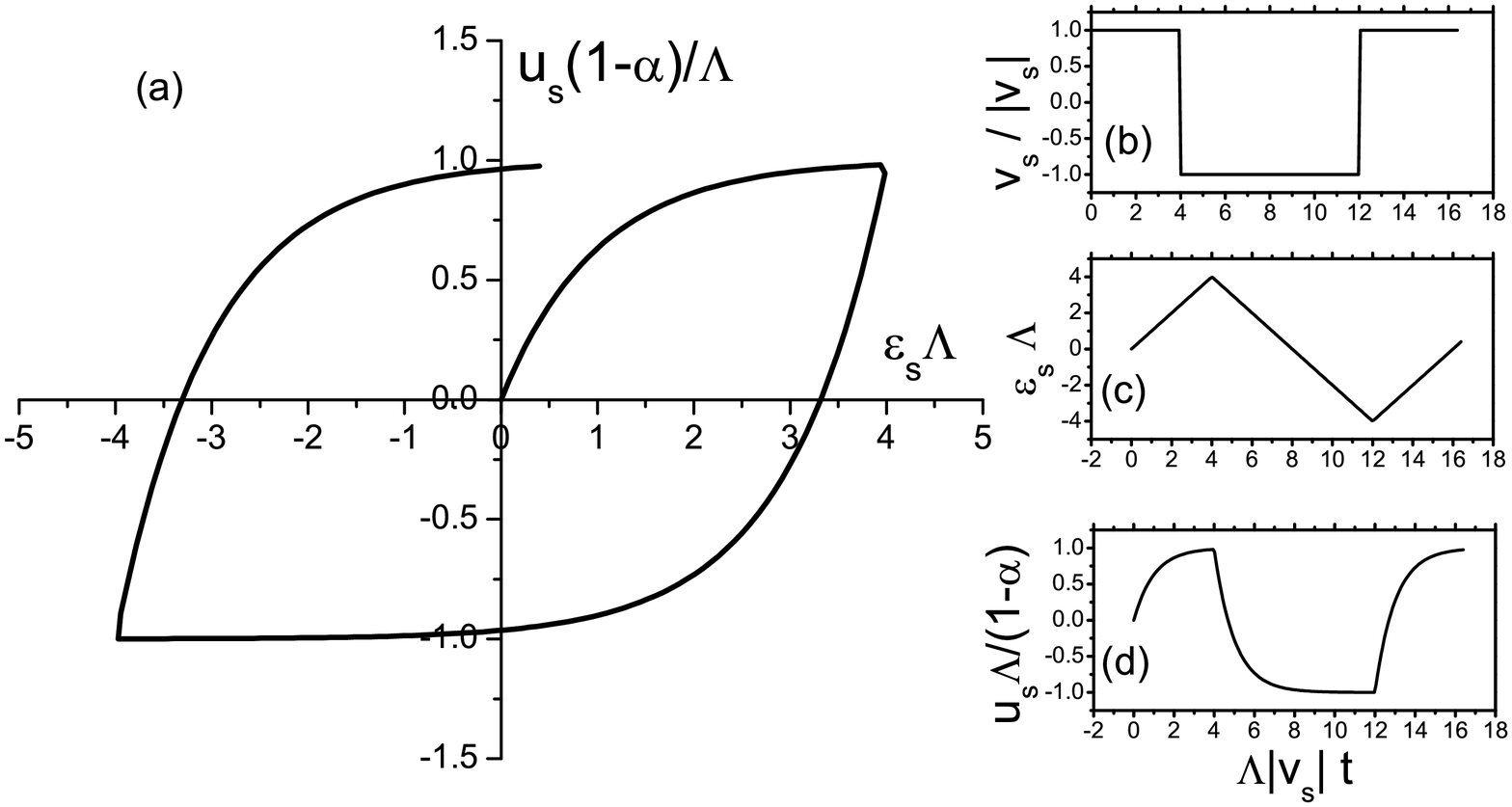}
\end{center}
\caption{\label{fig2} The hysteretic change of the shear stress ($\sim u_{s}$) with the strain, as accounted for by Eq~(\ref{3b-3}). The sign of the shear rate $v_{s}(t)$ is given in (b), the shear deformation $\varepsilon _{s}=\int_{0}^{t}v_{s}(t^{\prime })dt^{\prime }$ in
(c). Inset (d) is the the temporal evolution of $u_{s}$. } 
\end{figure} 

In this form, it is obvious that loading ($v_s=|v_s|>0$) and unloading ($v_s=- |v_s|<0$) have different slopes: $\partial_tu_s/v_s=(1-\alpha)\mp(\Lambda u_s/h)$.
This phenomenon is referred to as {\em
incremental nonlinearity} in soil mechanics, and the reason why no backtracking takes place under reversal of shear rate: Starting from isotropic stress, $u_s=0$, see Fig~\ref{fig2}, the gradient is at first $(1-\alpha)$, becoming smaller as $u_s$ grows, until it
is zero, in the stationary case $\partial_tu_s/v_s=0$. Unloading now, the
slope is $(1-\alpha)+(\Lambda u_s/h)$, steeper than it has ever been. It is again 
$(1-\alpha)$ for $u_s=0$, and vanishes for $u_s$ sufficiently negative, see Fig~\ref{fig2}. Same scenario holds for $\partial_t\Delta/v_s$. 

Clearly, only the stress $P,\sigma_s$ are measurable, not
$\Delta, u_s$. The former is calculated employing Eq~(\ref{sum stress hp}) when
the latter is given. The resultant expressions can be complicated
(especially if the pressure is held constant instead of the density), but the basic physics remains the same -- an illustration of why $u_{ij}$ is the better state variable.

In systematic studies employing discrete numerical simulation, Roux and coworkers have 
accumulated great knowledge about the mesoscopic physics on granular scales, see 
eg.~\cite{roux}. And they were especially able to distinguish between two types of strain, I 
and II, complete with two regimes in which either dominates. However, although  
type I strain may clearly be identified as our state variable $u_{ij}$, one needs to be aware that regime~I  is not necessarily {\em quasi-static}, or {quasi-elastic} as considered in Sec~\ref{quasi elastic motion}. The difference is: The relaxation term may be temporarily small at hypoplastic shear rates, say because $u_s$ or $\rho_{cp}-\rho$ are, see Eqs~(\ref{3b-3}, \ref{3b-4}), they do not stay small if one wanders in the variable space. At quasi-elastic rates, deformation are always free of dissipation.

\subsection{Stationary Elastic Solution, or the Critical State\label{critical state}} 
\begin{figure}[b] \begin{center}
\includegraphics[scale=0.27]{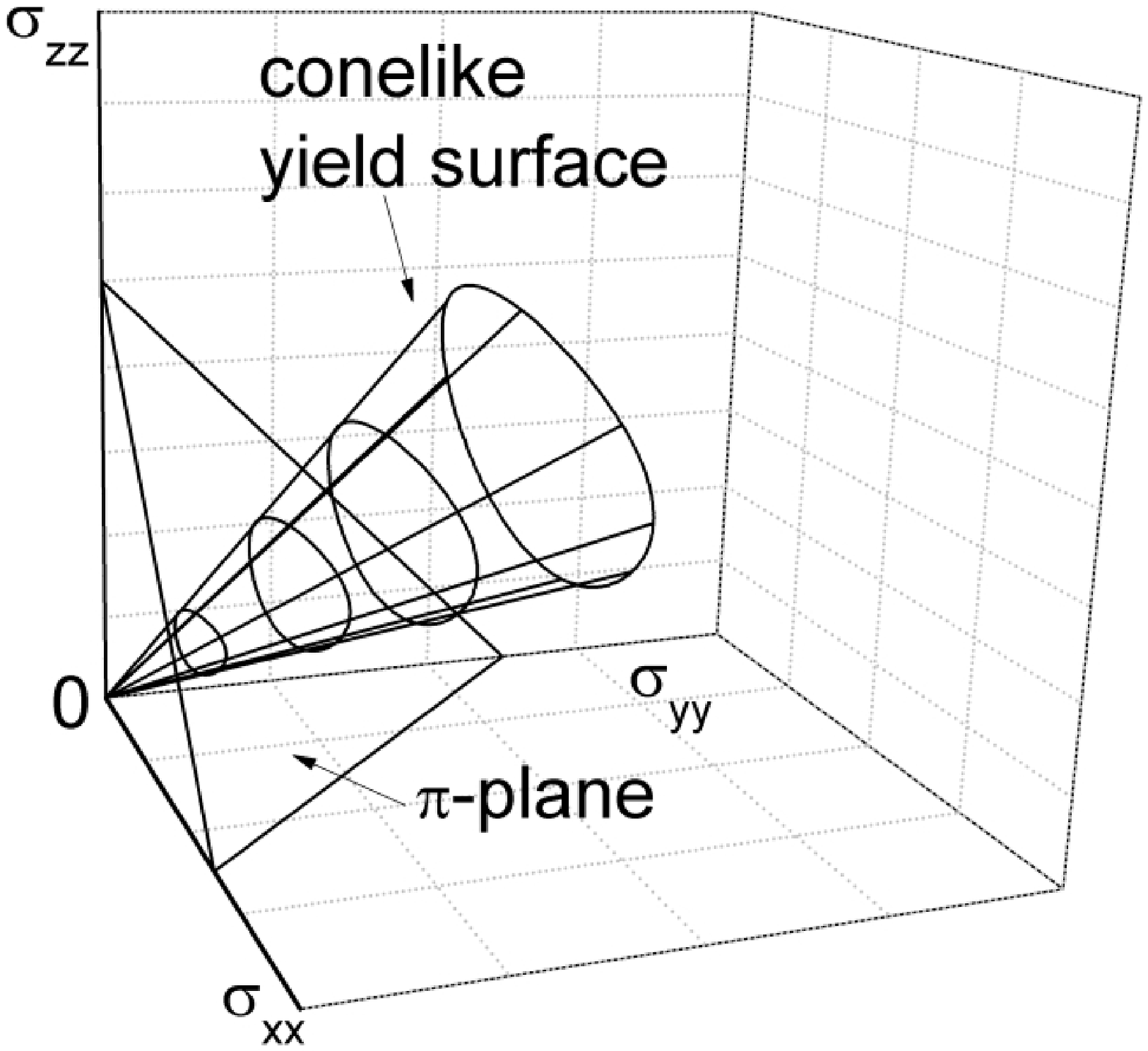}
\includegraphics[scale=0.2]{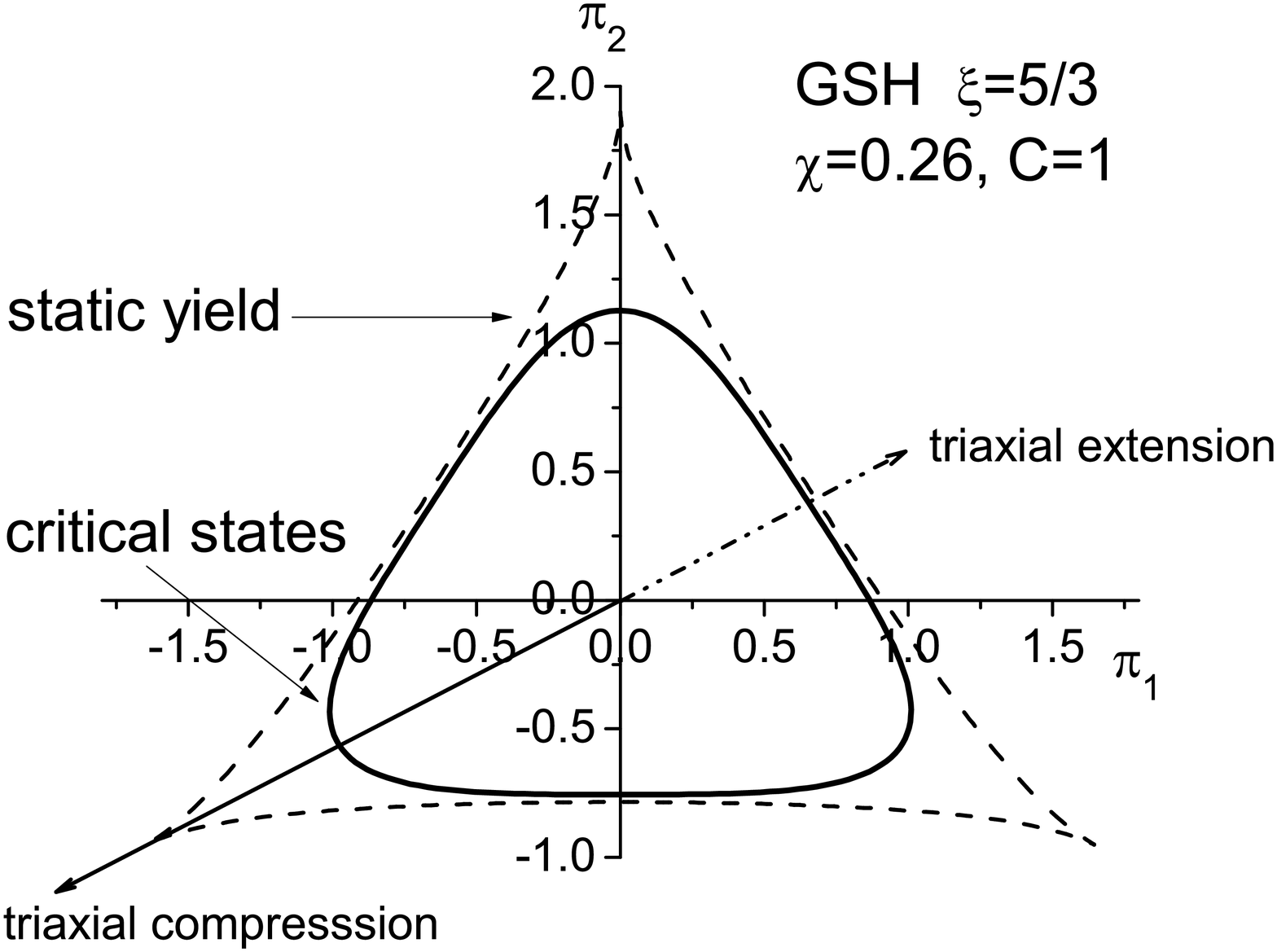}
\end{center}
\caption{\label{yield-cs-pai} Loci of static yield surface and the critical states, calculated employing the more general energy of Eq~(\ref{2b-6}). Left: in the space spanned by the three stress eigenvalues, $\sigma_1, \sigma_2, \sigma_3$; right: in the $\pi$-plane of constant pressure, $P\equiv\sigma_1+\sigma_2+\sigma_3$, where $\sqrt{2}\pi _{1} \equiv(\sigma _{3}-\sigma _{2})/P$, 
$\sqrt{6}\pi _{2} \equiv(2\sigma _{1}-\sigma _{2}-\sigma _{3})/P$. }
\end{figure}

When there is complete compensation of the shear rate $\sim v_s$ and the relaxation
$\sim T_g$, the stationary solution of Eqs~(\ref{3b-3}) for the elastic strain 
$u_{ij}$ holds. It is generally called the {\em critical state}, see~\cite{critState}, and may be considered ideally plastic, because a shear rate does not lead to a stress 
increase. Setting $\partial_t\Delta, \partial_tu_s=0$ in Eqs~(\ref{2c-7xx},\ref{2c-9xx}), we obtain the expressions, 
\begin{equation}\label{3b-3a} u_c=\frac{1-\alpha}{\lambda}\frac{v_s}{T_g}=\pm\frac{1-\alpha}{\Lambda},
\quad \frac{\Delta_c}{|u_c|}=\frac{\alpha_1}{\lambda_1}\frac{|v_s|}{T_g}=\frac{\alpha_1}{\Lambda_1}.
\end{equation} 
From Eq~(\ref{2c-8}), the collinearity of the critical strain and rate, $u^*_{ij}|_c/|u_c|=v^*_{ij}/|v_s|$, is easy to see. 
In the hypoplastic regime (for $h=1, \alpha=\bar\alpha,\alpha_1=\bar\alpha_1$),  $u_c,\Delta_c$ depend only on the density and is rate-independent. 
The critical stress is given by inserting $u_c,\Delta_c$ into Eqs~(\ref{sum stress hp}), 
\begin{eqnarray}\label{3b-3b} P^c_\Delta&\equiv&P_\Delta(\Delta_c,
u_c)=\sqrt\Delta_c({\cal B}\Delta_c+{\cal A} {u_c^2}/{2\Delta_c}),
\\\label{3b-3c}
\pi_c&\equiv&\pi_s(\Delta_c, u_c)=-2{\cal A}\sqrt\Delta_c\, u_c, 
\\\label{3b-3cA}
{P^c_\Delta}/{\pi_c}&=&({{\cal B}}/{2{\cal
A}}){\Delta_c}/{u_c}+{u_c}/{4\Delta_c}, 
\\\label{3b-4a}
P_c&=&(1-\bar\alpha)P^c_\Delta,\quad
\sigma_c=(1-\bar\alpha)\pi_c.
\end{eqnarray} 
The critical ratio $\sigma_c/P_c$ -- same as the Coulomb yield of Eq~(\ref{2b-3}) -- is also frequently associated with a friction angle. Since one is relevant for vanishing $T_g\sim v_s^2\to0$, while the other requires an elevated $T_g\sim |v_s|$, it is appropriate to
identify one as the static friction angle, and the other as the
dynamic one. The dynamic friction angle is always smaller than
the static one,  see Fig~\ref{yield-cs-pai}, because the critical state is elastic,
and must stay below Coulomb yield, 
\begin{equation}\label{3b-5} \Lambda_1/\bar\alpha_1<\sqrt{2{\cal B/A}}.
\end{equation}

Textbooks on soil mechanics frequently state that the friction angle is
essentially independent of the density -- although they do not, as a rule,
distinguish between the dynamic and the static one, cf. Sec~\ref{yield
surfaces}. We assume, for lack of more discriminating information, that both are. Therefore, we take $\alpha_1\sim(\rho_{cp}-\rho)$, because $\Lambda_1$ also does, see Eq~(\ref{3b-4}).
Quite generally, we note that accepting the density dependence of Eqs~(\ref{density-dependence}), we have $\Delta_c, u_c$ being monotonically increasing functions of $1/(\rho_{cp}-\rho)$. The same holds for $P_c,\sigma_c\sim{\cal B}$, though ${\cal B}$'s density dependence make the increase slightly faster.

\subsection{Constitutive Relations\label{conrel}}
As discussed in the introduction, granular dynamics is frequently modeled employing  the strategy of
{\em rational mechanics}, by postulating a function $\mathfrak{C}_{ij}$ -- of the  stress $\sigma _{ij}$, strain rate $v_{k\ell}$, and density $\rho $ --  such that the constitutive relation, ${\partial}_{t}\sigma _{ij}=\mathfrak{C}_{ij}(\sigma_{ij}, v_{k\ell}, \rho)$ holds (where ${\partial}_{t}$ is to be replaced by an appropriate objective derivative more generally). It forms, together with the continuity  equation $\partial _{t}\rho +\nabla _{i}\rho v_{i}=0$, momentum conservation, $\partial _{t}(\rho v_{i})+\nabla_{j}(\sigma _{ij}+\rho v_iv_j)=0$, a closed set of equations for $\sigma _{ij}$, the velocity $v_{i}$, and the density $\rho $ (or the void ratio $e$). Both hypoplasticity and barodesy considered below belong to this category. These models  yield, in circumstances where they hold, a realistic account of the complex elasto-plastic motion, providing us with highly condensed and intelligently organized empirical data. This enables us to validate {\sc gsh} and reduce the latitude in specifying  the energy and transport coefficients.  

At the same time, one needs to be aware of their drawbacks, especially the hidden ones. First of all is the apparent freedom in fixing $\mathfrak C$ -- constrained only by the data one considers, not by energy conservation or entropy production that were crucial in deriving {\sc gsh}.  This is what we believe the main reason why there are so many competing engineering models. As this liberty explodes when one includes gradient terms, most models refrain from the attempt to account for nonuniform situations, say elastic waves. Second, in dispensing with the variables $T_g$ and $u_{ij}$, and restricting the variables to $\sigma_{ij}, v_{k\ell}, \rho$, one reduces the model's range of validity and looses the benefit of $u_{ij}$'s simple behavior: First, the models of hypoplasticity and barodesy are valid only for  $T_g\sim|v_s|$, so a $T_g$ that is either too small or oscillates too fast will invalidate them.  Second, as the analytical solution of the approach to the critical state~\cite{critState} shows, considering $u_{ij}$ -- though it is not directly measurable -- is a highly simplifying intermediate step. The case for  $u_{ij}$ is even stronger, when considering proportional paths and the barodesy model, see below.  

\subsubsection{The Hypoplastic Model\label{Hypoplasticity}} 

The {\em hypoplastic model} starts from the rate-independent
constitutive relation, 
\begin{equation}\label{3b-1}
\partial_t\sigma_{ij}=H_{ijk\ell}v_{k\ell}+
\Lambda_{ij}\sqrt{v_s^2+\epsilon v_{\ell\ell}^2}, 
\end{equation} 
postulated by Kolymbas~\cite{kolymbas1}, where
$H_{ijk\ell},\Lambda_{ij},\epsilon$ are (fairly involved) functions of the
stress and packing fraction. Incremental nonlinearity as discussed in Sec~\ref{Load and Unload} is also part of the postulate. The simulated granular response is realistic for deformations at constant or slowly changing rates.
 
{\sc gsh} reduces to the hypoplastic model in the hypoplastic regime, for $T_g\sim|v_s|$, $\alpha =\bar \alpha,\alpha_1=\bar\alpha_1$,  $P_T, \eta_1T_g
v^0_{ij}\to0$. This is because $\sigma_{ij}=(1-\bar\alpha)\pi_{ij}$ of Eq~(\ref{sum stress hp}) is then, same as $\pi_{ij}$, a function of $u_{ij}, \rho$,  and we may write $\partial_t\sigma_{mn}=({\partial\sigma_{mn}}/{\partial u_{ij}})\partial_tu_{ij}+
({\partial\sigma_{mn}}/{\partial\rho})\partial_t\rho$.
Replacing $\partial_t\rho$ with the first of Eq~(\ref{2c-1}),
$\partial_tu_{ij}$ with Eq~(\ref{2c-8}), using Eq~(\ref{TgVs}) to eliminate $T_g$, we arrive at an equation with the same structure
as Eq~(\ref{3b-1}). Our derived result for $H_{ijk\ell},\Lambda_{ij}$ is
different from the postulated engineering expressions, and somewhat simpler, but they yield very
similar {\em response ellipses}, see~\cite{granL3}. (Response ellipses are
the strain increments as the response of the system, given unit stress
increments in all directions starting from an arbitrary point in the stress
space, or vice versa, stress increments as the response for unit strain increments.)

\subsubsection{Proportional Paths and Barodesy}

Barodesy is a very recent model, again proposed by Kolymbas~\cite{barodesy}. As compared to hypoplasticity, it is more modular and better organized, with different parts in $\mathfrak C_{ij}$ taking care of specific aspects of granular deformation, especially that of {\em proportional paths}. We take {\sc p}$\varepsilon${\sc p} and {\sc p}$\sigma${\sc p} to denote, respectively, proportional strain and stress path.  Their behavior is summed up by the Goldscheider rule ({\sc gr}): (1)~A {\sc p}$\varepsilon${\sc p} starting from the stress $\sigma_{ij}=0$ is associated with a  {\sc p}$\sigma${\sc p}.
(2)~A {\sc p}$\varepsilon${\sc p} starting from $\sigma_{ij}\not=0$
leads asymptotically to the corresponding  {\sc p}$\sigma${\sc p} obtained when starting at $\sigma_{ij}=0$.
(The initial value $\sigma_{ij}=0$ is a mathematical idealization, neither easily realized nor part of the empirical data that went into {\sc gr}. We take it  {\em cum grano salis}.) 

Explanation: Any constant strain rate $v_{ij}$ is a {\sc p}$\varepsilon${\sc p}. In the principal strain axes  $(\varepsilon_1,\varepsilon_2,\varepsilon_3)$, a constant $v_{ij}$ means the system moves with a constant rate along its direction, with $\varepsilon_1/\varepsilon_2=v_1/v_2,\, \varepsilon_2/\varepsilon_3=v_2/v_3$ independent of time. What {\sc gr} states is that there exists an associated stress path that is also proportional, also a straight line in the principal stress space, that there are pairs of strain and stress path which are linked,  and if the initial stress value is not on the right line, it will converge onto it.  

If {\sc gsh} is as claimed a broad-ranged theory on granular behavior, we should be able to  understand {\sc gr}  with it, which is indeed the case. Given any constant rate $v_{ij}$, the elastic strain will -- irrespective of its initial value, relax into the stationary state of Eqs~(\ref{2c-7xx},\ref{2c-9xx}),
\begin{equation}
\label{eq74}
u_c=\frac{1-\alpha}{\Lambda},\quad \frac{\Delta_c}{u_c}=\frac{\alpha _1}{\Lambda _{1}}+\frac{1-\alpha}{u_c\Lambda_1}\frac{v_{\ell\ell}}{v_s},
\end{equation}
with ${u_{ij}^{\ast }|_c}/{u_c}={v_{ij}^{\ast }}/{v_s}$.  Adding in the information from  Eqs~(\ref{2b-2a},\ref{2b-2b}), we also find
\begin{equation}
{\sigma_{ij}^{\ast }}/{\sigma_s}(\rho)={v_{ij}^{\ast }}/{v_s}.
\end{equation}
If the strain path is isochoric, with $v_{\ell\ell}=0$ and $\rho=$ const, both the deviatoric strain and stress are dots that remain stationary and do not walk down a path as time progresses. Clearly, these are simply the ideally plastic, stationary, critical state~\cite{critState}. If $v_{\ell\ell}\not=0$ with the density $\rho[t]$ changing accordingly, ${u_{ij}^{\ast }|_c}$ and ${\sigma_{ij}^{\ast }}$ will walk down a straight line along ${v_{ij}^{\ast }}/{v_s}$, with a velocity determined, respectively, by $u_c(\rho[t])$ and $\sigma_s(\rho[t])$. 

Given an initial strain deviating from that prescribed by Eq~(\ref{eq74}),  $u_0\not=u_c,\Delta_0\not=\Delta_c$, Eqs~(\ref{2c-7xx},\ref{2c-9xx}) clearly state that the deviation will exponentially relax, until they vanish -- ie., the strain and the associated stress will converge onto the prescribed line. All this is very well, but {\sc gr} states that it is the total stress that walks down a straight line. With
\begin{equation}
\pi_{ij}=P_\Delta(\rho)[\delta_{ij}+(\pi_s/P_\Delta)v_{ij}^*/v_s],
\end{equation}
this fact clearly hinges on $(\pi_s/P_\Delta)$ -- a function of $\Delta/u_s$ [see Eq~(\ref{2b-3b})] --  not depending on the density. As long as $v_{\ell\ell}\ll v_s$, we have  $\Delta_c/u_c\approx{\alpha _1}/{\Lambda _{1}}$, a combination that we did assume is density independent, see Eq~(\ref{density-dependence}), partially in anticipation of the fact that the friction angle of the critical state, a function of $(\pi_s/P)$, is independent of the density. And $v_{\ell\ell}/v_s$ must indeed remain small to avoid hitting either $\rho_{cp}$ or $\rho_{lp}$ too quickly.

In~\cite{GSH&Barodesy}, the results of {\sc gsh} are compared to that of barodesy, with mostly quantitative agreement. (The energy of Eq~(\ref{2b-6}) was employed there. So the results are more realistic.) When looking at $\mathfrak C_{ij}$, it is easy to grasp that  the construction of a constitutive relation is only for someone with vast experience about granular media. That we could substitute this deep knowledge with the notions of {\em variable transient elasticity}, giving rise to a theory just as capable of accounting for elasto-plastic motion, is eye-opening. It suggests that sand, in its qualitative behavior, may be after all neither overly complicated, nor such a rebel against general principles.


\subsection{Elastic Waves\label{elastic waves}}
That elastic waves propagate in granular media~\cite{jia1999,jia2004} is an important fact, because it is an unambiguous proof that granular media possess an 
elastic regime, and behave as elastic media in certain parameter ranges. Experimental exploration of the elastic to plastic transition would be equally crucial, and elastic waves remain a useful tool for this purpose.  

There is a wide-spread believe that small, quasi-static increments from any 
equilibrium stress state is elastic, but large ones are plastic.  As discussed in 
Sec~\ref{3regimes}, this assumption is illogical, because a large increment is 
the sum of small ones. In {\sc gsh}, the parameter that sets the boundary between elastic and plastic regime is the granular temperature $T_g$. We have 
quasi-elastic regime for vanishing $T_g\sim v_s^2$, and the 
hypoplastic one for elevated $T_g\sim v_s$. 

A perturbation in the elastic strain or stress propagate as a wave only in the quasi-elastic regime, while it diffuses in the hypoplastic one. 
More specifically, we shall derive a telegraph equation from {\sc gsh}, with a quantity $\sim T_g$ taking on the role of the electric resistance~\cite{zhang2012}. It defines a characteristic  frequency $\omega_0\sim T_g$, such that elastic perturbations of the frequency $\omega$ diffuse for $\omega\ll\omega_0$, and propagate for $\omega\gg\omega_0$.  In the quasi-elastic regime, $\omega_0\to0$, and all perturbations propagate. In the hypoplastic regime, when $T_g$ gets elevated, so does $\omega_0$, pushing the propagating range to ever higher frequencies. Eventually, the associated wave length become comparable to the granular diameter,  exceeding  {\sc gsh}'s range of validity.

To derive the telegraph equation, we start with two basic equations of {\sc gsh}, Eqs~(\ref{2c-8},\ref{2c-1}),
\begin{eqnarray} 
\rho\partial_tv_i-(1-\alpha)\nabla_mK_{imkl}u^*_{kl}=0, \quad
\partial_tu^*_{ij}-(1-\alpha)v^0_{ij}=-\lambda T_gu^*_{ij},  \label{ew2}
\end{eqnarray}
where $K_{imkl}\equiv-\partial^2w/\partial u_{im}\partial u_{kl}$. (For simplicity, we concentrate on shear waves, assuming $v_{\ell\ell}\equiv0 $.) 
For $T_g\to0$, both plastic terms $\lambda T_gu^*_{ij}$ and $\alpha\sim T_g$ are  negligibly small, such that these two equations reproduce conventional elasticity theory. The variation of wave velocities $c$ with stress and density is then easily calculated, because $c^2$ is given by the eigenvalues of the matrix $K_{imnj}q_mq_n/(\rho q^2)$  ($q_m$ is the wave vector). The result~\cite{ge4} agrees well with observations~\cite{jia2009}.

There are two ways to crank up $T_g$. First is to introduce an ambient temperature, such as by tapping or a remote shear band, second is to increase the amplitude of the wave mode, because its own shear rate also creates $T_g$. The granular temperature has a characteristic time $\tau_T=b\rho/\gamma_1$, see Eq~(\ref{sum T_g}), that is of order $10^{-3}$~s in dense media. For simplicity, we assume that the  wave mode's  frequency is much larger than $1/\tau_T$ , such that $T_g$ and $\alpha(T_g)$ are essentially constant. This implies
\begin{eqnarray}\label{ew3}
(\partial^2_t+\lambda T_g\partial_t)\,u^*_{ij}={\textstyle\frac12}(1-\alpha)^2\times
\nabla_m[K_{imkl}\nabla_ju^*_{kl}+K_{jmkl}\nabla_iu^*_{kl}].
\nonumber
\end{eqnarray}
Concentrating on one wave mode propagating along $x$, with $c_{\rm qs}$ the quasi-elastic velocity and $\bar u\sim e^{iqx-i\omega t}$ the amplitude of the associated eigenvector, we obtain the telegraph equation, 
\begin{equation}\label{ew4}
(\partial^2_t+\lambda T_g\partial_t)\,\bar u= (1-\alpha)^2c_{\rm qs}^2\nabla_x^2\,\bar u
\equiv c^2\nabla_x^2\,\bar u.
\end{equation}
(The coefficient $\alpha$ accounts for the fact that granular contacts soften with $T_g$, and the effective elastic stiffness decreases by $(1-\alpha)^2$. In the language of electromagnetism, 
 $(1-\alpha)^{-2}$  is a dielectric permeability.)
Inserting $\bar u\sim e^{iqx-i\omega t}$ into Eq~(\ref{ew4}), we find
\begin{equation}
c^2 q^2={\omega^2+i\omega\lambda T_g},
\end{equation}
implying diffusion for the low frequency limit, $\omega\ll\lambda T_g$,
\begin{equation}
q\approx\pm\frac{\sqrt{\omega\lambda T_g}}c \, \frac{1+i}{\sqrt2},
\end{equation}
and propagation for the high-frequency limit, $\omega\gg\lambda T_g$,
\begin{eqnarray}
q\approx\pm\frac\omega c\left(1+i\,\frac{\lambda T_g}{2\omega}
\right),
\quad
\bar u\sim\exp{\left[-i\omega\left(t\mp\frac xc\right) 
t\mp x\frac{\lambda T_g}{2 c}\right]}.
\end{eqnarray}
The first term in the square bracket accounts for wave propagation, the second a decay length $2c/\lambda T_g$, which is frequency-independent  if $T_g$ is an ambient temperature. If  $T_g$ is produced by the elastic wave itself,  it varies between $T_g\sim v_s^2\sim\omega^2q^2\sim\omega^4$ and $T_g\sim v_s\sim\omega^2$ depending on the amplitude, and the decay length is strongly frequency dependent.
     
A brief wave pulse, arbitrarily strong, can always propagate through granular media if its duration is too brief to excite sufficient $T_g$ for the system to enter the hypoplastic regime. The duration must be much smaller than the characteristic time $b\rho/\gamma$ of $T_g$, see Eq~(\ref{sum T_g}). 
        
\section{Conclusions}
{\sc gsh} is derived employing the hydrodynamic approach, starting from two assumptions about granular media's basic physics: {\it variable transient elasticity} and {\it two-stage irreversibility}. Because of the many constraints this approach provides, {\sc gsh} should be a valid broad-ranged theory, from statics to fast dense flow, if these assumptions are  appropriate.
{\it Variable transient elasticity} prescribes relaxation of the elastic stress with a rate $\sim T_g$, with full elasticity restored at equilibrium, for $T_g=0$. And indeed,  this leads to results in several static geometries, including silos, sand pile and a layer subject to a point load, that agree well with data. Same is true for the incremental stress-strain relation and velocities of sound waves, both calculated setting $T_g=0$. Finally,  we conclude that the quasi-static motion in {\sc gsh} is quasi-elastic, a visit of consecutive elastic states at  $T_g=0$.

{\it Two-stage irreversibility} defines what the granular temperature $T_g$ is, and provides a  relation between  $T_g$ and the shear rate $\dot\gamma$, in the stationary state in which $T_g$ is a constant. Given by transport coefficients (the functional dependence of which is an input),  this relation is not completely fixed. Assuming the simplest dependence, we have $T_g\sim\dot\gamma$, implying more stress relaxation the faster the shear motion is. This is the physics of rate-independence: Because the same shear motion also deforms the grains and builds up the elastic stress, a motion double as fast with twice the relaxation rate leads to the same stress. At given $\dot\gamma$, the elastic stress relaxes toward its stationary solution, in which the elastic deformation and dissipative relaxation cancel, implying  a constant stress. This is  the critical state. 

However, a rate-independent ratio between the elastic deformation and dissipative relaxation means that the latter does not vanish more quickly than the former for $\dot\gamma\to0$,  implying the complete lack of quasi-static motion, which is, of course, a fairly general phenomenon. Although there is as yet not much direct experimental evidence for its existence, we note that it may be restored by changing the transport coefficients slightly, such that  $T_g\sim\dot\gamma$ goes smoothly over to  $T_g\sim\dot\gamma^2$ for  $\dot\gamma\to0$, because a quadratically small $T_g$ implies a dissipation that vanishes more quickly than the  elastic motion. 
As a result, {\sc gsh} has three rate regimes, given by: 
\begin{itemize}
\item The {\em quasi-elastic regime} of vanishing shear rates $\dot\gamma$, possibly below $10^{-5}$~s$^{-1}$, with a quadratically vanishing granular temperature, $T_g\sim\dot\gamma^2\to0$. The stress $\sigma_{ij}=\pi_{ij}$  is purely elastic, as given by Eq~(\ref{1-1}).  
This regime is admittedly difficult to observe. Some possible reasons are discussed in Sec~\ref{aeipt}, with suggestions in~\ref{soft springs} on how to overcome them. 

\item The {\em hypoplastic regime} of low shear rates, possibly between $10^{-3}$ and 1~s$^{-1}$, where the engineering theories such as the hypoplastic model~\cite{kolymbas1} holds. The stress $\sigma_{ij}=(1-\alpha)\pi_{ij}$ is still elastic, but softer 
by the factor $1-\alpha$, typically between 0.2 and 0.3. Granular temperature is more elevated, allowing stress relaxation. Rate-independence is a result of  $T_g\sim\dot\gamma$.  
As we have seen, three simple scalar equations are sufficient to account for general elasto-plastic motion, including especially load-unload behavior, Sec~\ref{Load and Unload}; and the approach to the critical state, Sec~\ref{critical state}. They were also used for a successful comparison to the hypoplastic and barodesy model, Sec~\ref{conrel}; and for the damping of elastic waves, Sec~\ref{elastic waves}.

This regime is frequently termed the quasi-static one, because it lacks inertial effects, is rate-independent, and the even slower quasi-elastic regime is hard to 
observe. We note in Sec~\ref{yield surfaces} and~\ref{3regimes} that the hypoplastic regime, characterized by stress relaxation,  is strongly dissipative.

\item The {\em rapid flow regime}, for shear rates well above 1 s$^{-1}$. We still have $T_g\sim\dot\gamma$, but it is no longer small. Therefore, the $T_g$-generated, seismic pressure $P_T\sim T_g^2\sim\dot\gamma^2$ and the viscous shear stress $\sigma_s\sim T_g\dot\gamma\sim\dot\gamma^2$ become significant and compete with the elastic contribution $\pi_{ij}$. This is where the MIDI model and Bagnold flow hold. As both the pressure and the shear stress may be written as $e_1+e_2\dot\gamma^2$, where $e_1$ is the elastic, and $e_2$ the seismic, or viscous, contributions,  we have a quadratic dependence of the Bagnold flow for $e_2\dot\gamma^2\gg e_1$, and hypoplastic rate-independence for  $e_2\dot\gamma^2\ll e_1$. 
This rate regime has already been considered in~\cite{p&g2013}.
\end{itemize}
Finally, a summary of some frequently used  quantities, for which  physics and engineering textbooks employ rather different notations,  
first a general tensor, say $v_{ij}$. We take $v_{\ell\ell}$ as its trace, $v_{ij}^*$ as its traceless part, with $v_s\equiv v_{ij}^*v_{ij}^*$  as the second invariant. 
Then stresses and strains: total or Cauchy stress: $\sigma_{ij}$,  the elastic stress: $\pi_{ij}$, with $\sigma_s,\pi_s$ as defined above. The elastic strain: $u_{ij}$, with $\Delta\equiv- u_{\ell\ell}$ and $u_s\equiv u_{ij}^*u_{ij}^*$, is defined around Eq~(\ref{1-1}). The strain rate (frequently denoted as $\dot\epsilon_{ij}$) is taken as $v_{ij}\equiv\frac12(\nabla_iv_j+\nabla_jv_i)$, and the scalar shear rate  $v_s\equiv v_{ij}^*v_{ij}^*$ (or interchangeably, $\dot\gamma$).

The granular temperature is $T_g$, note the energy is $\sim T_g^2$, see Sec~\ref{significanceT_g}.

\noindent{\bf Acknowledgment:} We thank Itai Einaf for a critical reading of the manuscript and many help- and  insightful comments.


\begin{thebibliography}{99} 

\bibitem{LL6}
L.~D. Landau and E.~M. Lifshitz.
\newblock {\em Fluid Mechanics}.
\newblock Butterworth-Heinemann, 1987.

\bibitem{Khal}
I.~M. Khalatnikov.
\newblock {\em Introduction to the Theory of Superfluidity}.
\newblock Benjamin, New York, 1965.

\bibitem{deGennes}
P.G. de~Gennes and J.~Prost.
\newblock {\em The Physics of Liquid Crystals}.
\newblock Clarendon Press, Oxford, 1993.

\bibitem{schofield}
P.~Wroth A.~Schofield.
\newblock {\em Critical State Soil Mechanics}.
\newblock McGraw-Hill, London, 1968.

\bibitem{nedderman}
R.M. Nedderman.
\newblock {\em Statics and Kinematics of Granular Materials}.
\newblock Cambridge University Press, 1992.

\bibitem{wood1990}
D.~M. Wood.
\newblock {\em Soil Behaviour and Critical State Soil Mechanics}.
\newblock Cambridge University Press, 1990.

\bibitem{kolymbas1}
D.~Kolymbas.
\newblock {\em Introduction to Hypoplasticity}.
\newblock Balkema, Rotterdam, 2000.

\bibitem{kolymbas2}
W.~Wu and D.~Kolymbas.
\newblock {\em Constitutive Modelling of Granular Materials}.
\newblock Springer, Berlin, 2000.

\bibitem{gudehus2010}
G.~Gudehus.
\newblock {\em Physical Soil Mechanics}.
\newblock Springer SPIN, 2010.

\bibitem{hutter2007}
S.P. Pudasaini and K.~Hutter.
\newblock {\em Avalanche Dynamics}.
\newblock Springer, 2007.
\bibitem{hydro-1}S. R. de Groot and P. Masur,
{\it Non-Equilibrium Thermodynamics}, (Dover, New York 1984).

\bibitem{hydro-2}D. Forster, Hydrodynamic Fluctuations, Broken Symmetry and
Correlation Functions (Benjamin, New York, 1975).

\bibitem{liqCryst-1} P.G. de Gennes and J. Prost, {\em The Physics of
Liquid Crystals} (Clarendon Press, Oxford 1993). 

\bibitem{liqCryst-2} P.C.
Martin, O. Parodi, and P.S. Pershan, {\it Unified Hydrodynamic Theory for
Crystals, Liquid Crystals, and Normal Fluids}, Phys. Rev. A 6, 2401 (1972).

\bibitem{liqCryst-3} T.C. Lubensky, {\it Hydrodynamics of Cholesteric
Liquid Crystals}, Phys. Rev. A 6, 452 (1972). 

\bibitem{liqCryst-4} M. Liu,
{\it Hydrodynamic Theory near the Nematic Smectic-A Transition}, Phys. Rev.
{\bf A 19}, 2090 (1979); 

\bibitem{liqCryst-5} M. Liu, {\it Hydrodynamic
theory of biaxial nematics}, Phys. Rev. {\bf A 24}, 2720 (1981).
\bibitem{liqCryst-6} M. Liu, {\it Maxwell equations in nematic liquid
crystals}, Phys. Rev. {\bf E 50}, 2925, (1994). 

\bibitem{liqCryst-7} H.
Pleiner and H.R. Brand, in {\it Pattern Formation in Liquid Crystals},
edited by A. Buka and L. Kramer (Springer, New York, 1996).

\bibitem{he3-1} R. Graham, {\it Hydrodynamics of 3He in Anisotropic A
Phase}, Phys. Rev. Lett. {\bf 33}, 1431 (1974). 

\bibitem{he3-2} R. Graham
and H. Pleiner, {\it Spin Hydrodynamics of 3He in the Anisotropic A Phase},
Phys. Rev. Lett. {\bf 34}, 792 (1975). 

\bibitem{he3-3} M. Liu, {\it
Hydrodynamics of $^3$He near the A-Transition,} Phys. Rev. Lett. {\bf 35},
1577 (1975). 

\bibitem{he3-4} M. Liu and M.C. Cross, {\it Broken Spin-Orbit
Symmetry in Superfluid $^3$He and the B-Phase Dynamics,} Phys. Rev. Lett.
{\bf 41}, 250 (1978). 

\bibitem{he3-5} M. Liu and M.C. Cross, {\it Gauge
Wheel of Superfluid $^3$He,} Phys. Rev. Lett. {\bf 43}, 296 (1979).

\bibitem{he3-6} M. Liu, {\it Relative Broken Symmetry and the Dynamics of
the $A_1$-Phase,} Phys. Rev. Lett. {\bf 43}, 1740 (1979).

\bibitem{SC-1} M. Liu, {\it Rotating Superconductors and the
Frame-independent London Equations,} Phys. Rev. Lett. {\bf 81}, 3223,
(1998). 

\bibitem{SC-2} Jiang Y.M. and M. Liu, {\it Rotating Superconductors
and the London Moment: Thermodynamics versus Microscopics,} Phys. Rev. {\bf
B 6}, 184506, (2001). 

\bibitem{SC-3} M.~Liu, {\em Superconducting
Hydrodynamics and the Higgs Analogy,} J. Low Temp. Phys. 126, 911, (2002)

\bibitem{hymax-1} K. Henjes and M. Liu, {\it Hydrodynamics of Polarizable
Liquids,} Ann. Phys. {\bf 223}, 243 (1993). 

\bibitem{hymax-2} M. Liu, {\it
Hydrodynamic Theory of Electromagnetic Fields in Continuous Media,} Phys.
Rev. Lett. {\bf 70}, 3580 (1993). 

\bibitem{hymax-3} {\it Mario Liu
replies,} Phys. Rev. Lett. {\bf 74}, 1884, (1995). 

\bibitem{hymax-4} Y.M.
Jiang and M. Liu, {\it Dynamics of Dispersive and Nonlinear Media,} Phys.
Rev. Lett. {\bf 77}, 1043, (1996).

\bibitem{FF-1} M.I. Shliomis, {\em Magnetic Fluids}, Sov. Phys. Usp. 17,
153 (1974). 

\bibitem{FF-2} R.E. Rosensweig, {\em Ferrohydrodynamics},
(Dover, New York 1997). 

\bibitem{FF-3} M. Liu, {\it Fluiddynamics of
Colloidal Magnetic and Electric Liquid,} Phys. Rev. Lett. {\bf 74}, 4535
(1995). 

\bibitem{FF-4} M. Liu, {\it Off-Equilibrium, Static Fields in
Dielectric Ferrofluids,} Phys. Rev. Lett. {\bf 80}, 2937, (1998).

\bibitem{FF-5} M. Liu, {\it Electromagnetic Fields in Ferrofluids}, Phys.
Rev. {\bf E 59}, 3669, (1999). 

\bibitem{FF-6} H.W.~M\"{u}ller and M.~Liu,
{\it Structure of Ferro-Fluiddynamics,} Phys. Rev. {\bf E 64}, 061405
(2001). 

\bibitem{FF-7} H.W. M\"{u}ller and M. Liu, {\em Shear Excited Sound
in Magnetic Fluid}, Phys. Rev. Lett. {\bf 89}, 67201, (2002).

\bibitem{FF-8} O. M\"{u}ller, D. Hahn and M. Liu, {\em Non-Newtonian
behaviour in ferrofluids and magnetization relaxation,} J. Phys.: Condens.
Matter 18, 2623, (2006). 

\bibitem{FF-9} S. Mahle, P. Ilg and M. Liu, {\em
Hydrodynamic theory of polydisperse chain-forming ferrofluids,} Phys. Rev.
{\bf E 77}, 016305 (2008).

\bibitem{polymer-1} H. Temmen, H. Pleiner, M. Liu and H.R. Brand, {\it
Convective Nonlinearity in Non-Newtonian Fluids,} Phys. Rev. Lett. {\bf
84}, 3228 (2000). 

\bibitem{polymer-2}H. Temmen, H. Pleiner, M. Liu and H.R.
Brand,{\it Temmen et al. reply}, Phys. Rev. Lett. {\bf 86}, 745 (2001).

\bibitem{polymer-3} H. Pleiner, M. Liu and H.R. Brand, {\it Nonlinear Fluid
Dynamics Description of non-Newtonian Fluids}, {Rheologica Acta} {\bf 43},
502 (2004). 

\bibitem{polymer-4}O. M\"{u}ller, {\em Die Hydrodynamische
Theorie Polymerer Fluide}, PhD Thesis University T\"{u}bingen (2006).

\bibitem{midi}
GDR MiDi.
\newblock On dense granular flows.
\newblock {\em The European Physical Journal E}, 14(4):341--365 (2004).


\bibitem{p&g2013}Yimin Jiang and Mario Liu, 
{\it AIP Conf. Proc.} {\bf 1542}, pp. 52 (2013); doi: http://dx.doi.org/10.1063/1.4811867 
\bibitem{LL5}L.D. Landau, and  E.M. Lifshitz, {\it Statistical Physics}, Butterworth-Heinemann, 1980

\bibitem{2-p&g2013}Yanpei Chen, Meiying Hou, Pierre Evesque, Yimin Jiang, and Mario Liu: {\it AIP Conf. Proc.} {\bf 1542}, 791 (2013); doi: 10.1063/1.4812050



\bibitem{luding2009}
Stefan Luding.
\newblock Towards dense, realistic granular media in 2d.
\newblock {\em Nonlinearity}, 22:101--146, 2009.

\bibitem{Bocquet}  
L.~Bocquet, W.~Losert, D.~Schalk, T.~C. Lubensky, and J.~P. Gollub.
\newblock Granular shear flow dynamics and forces: Experiment and continuum
  theory.
\newblock {\em Phys. Rev. E}, 65(1):011307, Dec 2001.


\bibitem{Houlsby}
G.~T. Houlsby and A.~M. Puzrin.
\newblock {\em Principles of Hyperplasticity}.
\newblock Springer (2006).


\bibitem{Houlsby2}
I.~F. Collins and G.~T. Houlsby.
\newblock Application of thermomechanical principles to the modelling of
  geotechnical materials.
\newblock {\em Proc. R. Soc. Lond. A}, 453:1975--2001, 1997.

\bibitem{rubin} 
M.B: Rubin, Physical reasons for abandoning plastic deformation measures in plasticity 
and viscoplasticity theory. Arch. Mech. 53 (4–5), 519–553 (2001).

\bibitem{granR2}
Y.~Jiang and M.~Liu. Granular solid hydrodynamics.  {\em Granular Matter}, 11:139, May 2009.\\ Free download: {www.springerlink.com/content/a8016874j8868u8r/fulltext} 

\bibitem{granR3}
Y.~Jiang and M.~Liu.
\newblock The physics of granular mechanics.
\newblock In D.~Kolymbas and G.~Viggiani, editors, {\em Mechanics of Natural
  Solids}, pages 27--46. Springer, 2009.


\bibitem{gudehus-jl}
G.~Gudehus, Y.M. Jiang, and M.~Liu.
\newblock Seismo- and thermodynnamics of granular solids.
\newblock {\em Granular Matter}, 1304:319--340, 2011.

\bibitem{kadanoff}
L.~P. Kadanoff.
\newblock Built upon sand: Theoretical ideas inspired by granular flows.
\newblock {\em Reviews of Modern Physics}, 71 (1):435 -- 444 (1999).

\bibitem{onsager}V. Garzo, J. M. Montanero, and J. W. Dufty, Phys. Fluids 18, 083305 (2006). It is not clear to us where the discrepancy arises, perhaps because only the production of granular entropy is considered, not that of the true entropy. There is a two-step dissipation in granular media: macroscopic energy $\to$ granular heat $\to$ true heat. Only when the second step is included is the description complete. 




\bibitem{denseFlow}
Stefan~Mahle, Yimin~Jiang and Mario~Liu.
\newblock Granular solid hydrodynamics: Dense flow, fluidization and jamming.
\newblock {\em arXiv:1010.5350v1 [cond-mat.soft]}, 2010.


\bibitem{ge1}
D.~O. Krimer, M.~Pfitzner, K.~Br\''auer, Y.~Jiang, and M.~Liu.
\newblock Granular elasticity: General considerations and the stress dip in
  sand piles.
\newblock {\em Phys. Rev. E)}, 74(6):061310, 2006.

\bibitem{ge2}
K.~Br\"auer, M.~Pfitzner, D.~O. Krimer, M.~Mayer, Y.~Jiang, and M.~Liu.
\newblock Granular elasticity: Stress distributions in silos and under point
  loads.
\newblock {\em Phys. Rev. E (Statistical, Nonlinear, and Soft Matter Physics)},
  74(6):061311, 2006.

\bibitem{kuwano2002}
R.~Kuwano and R.~J. Jardine.
\newblock On the applicability of cross-anisotropic elasticity to granular
  materials at very small strains.
\newblock {\em Geotechnique}, 52(10):727--749, Dec 2002.

\bibitem{ge3}
Y.M.~Jiang and M.~Liu.
\newblock Incremental stress-strain relation from granular elasticity:
  Comparison to experiments.
\newblock {\em Phys. Rev. E (Statistical, Nonlinear, and Soft Matter Physics)},
  77(2):021306, 2008.
  
\bibitem{jia2009}
Y.~Khidas and X.~Jia.
\newblock Anisotropic nonlinear elasticity in a spherical-bead pack: Influence
  of the fabric anisotropy.
\newblock {\em Phys. Rev. E}, 81:021303, Feb. 2010.


\bibitem{ge4}
M.~Mayer and M.~Liu.
\newblock Propagation of elastic waves in granular solid hydrodynamics.
\newblock {\em Phys. Rev. E}, 82:042301, 2010.

\bibitem{hardin}
B.O. Hardin and F.E. Richart.
\newblock Elastic wave velocities in granular soils.
\newblock {\em J. Soil Mech. Found. Div. ASCE}, 89: SM1:33--65, 1963.
\bibitem{einaf}M.B. Rubin and I. Einav,  A large deformation breakage model of granular materials including porosity and inelastic distortional deformation rate. {\it International Journal of Engineering Science}, {\bf 49} 1151–1169  (2011).
\bibitem{lade-duncan}
P.V. Lade and J.M. Duncan.
\newblock Elastoplastic stress-strain theory for cohesionless soil.
\newblock {\em Proc. ASCE, JGTD,}, 101:N0 GT10, 1975.

\bibitem{matsuoka}
H.~Matsuoka and T.~Nakai.
\newblock Stress-strain relationship of soil based on the smp.
\newblock {\em Proc. 9th ICSMFE}, specialty session 9:153--163, 1977.

\bibitem{3inv} Y.M. Jiang, H.P. Zheng, Z. Peng, L.P. Fu, S.X. Song, Q.C. Sun, M. Mayer, and M. Liu, \newblock Expression for the granular elastic energy.
\newblock  Phys. Rev. E {\bf 85}, 051304 (2012)    


\bibitem{granL3}
Y.~Jiang and M.~Liu.
\newblock From elasticity to hypoplasticity: Dynamics of granular solids.
\newblock {\em Phys. Rev. Lett.}, 99(10):105501, 2007.


\bibitem{roux} J.-N. Roux.  How granular materials deform in quasistatic conditions  AIP Conf. Proc. 1227, pp. 260-270; doi:http://dx.doi.org/10.1063/1.3435396; The nature of quasi-static deformation in granular materials. {\em arXiv:0901.2305v1 [cond-mat.soft]}, 2009;

\bibitem{critState}
Stefan Mahle, Yimin Jiang, and Mario Liu.
\newblock The critical state and the steady-state solution in granular solid
  hydrodynamics.
\newblock {\em arXiv:1006.5131v3 [physics.geo-ph]}, 2010.



\bibitem{barodesy} Kolymbas D. Barodesy: a new constitutive frame for soils. Geotechnique Letters 2, 17–23,  (2012), http://dx.doi.org/10.1680/geolett.12.00004; 
Barodesy: A new hypoplastic approach. International Journal for Numerical and Analytical Methods in Geomechanics (2011). doi:10.1002/nag.1051;
Sand as an archetypical natural solid. In Mechanics of Natural Solids, Kolymbas D, Viggiani G (eds.). Springer: Berlin, (2009); 1–26; 

\bibitem{GSH&Barodesy} Yimin Jiang, and Mario Liu. Proportional Path, Barodesy, and Granular Solid Hydrodynamics. Preprint

\bibitem{jia1999}
X.~Jia, C.~Caroli, and B.~Velicky.
\newblock Ultrasound propagation in externally stressed granular media.
\newblock {\em Phys. Rev. Lett.}, 82(9):1863--1866, Mar 1999.

\bibitem{jia2004}
X.~Jia.
\newblock Codalike multiple scattering of elastic waves in dense granular
  media.
\newblock {\em Phys. Rev. Lett.}, 93(15):154303, Oct 2004.

\bibitem{zhang2012}
Q.~Zhang, Y.C. Li, M.Y. Hou, Y.M. Jiang, and M.~Liu.
\newblock Elastic waves in the presence of a granular shear band formed by
  direct shear.
\newblock {\em Phys. Rev. E}, 85:031306, 2012.

\end{thebibliography}
\end{document}